\documentclass[11pt,twoside,a4paper,cmspaper,final,collab]{cms-tdr}
\def\svnVersion{6de24d2}\def\svnDate{2026/08/24}\def\cmsCernNoTag{CERN-EP-2026-001}\def\cmsCernDate{\today}\def\cmsMessage{Published in the Journal of High Energy Physics as \href{http://dx.doi.org/10.1007/JHEP08(2026)050}{\doi{10.1007/JHEP08(2026)050}.}}
\begin{document}\cmsNoteHeader{B2G-23-007}

\providecommand{\PX}{{\HepParticle{X}{}{}}\xspace}
\newcommand{\PS}{{\HepParticle{S}{}{}}\xspace}
\newcommand{\PY}{{\HepParticle{Y}{}{}}\xspace}
\newcommand{\PG}{{\HepParticle{G}{}{}}\xspace}
\newcommand{\PA}{{\HepParticle{A}{}{}}\xspace}
\providecommand{\Pa}{{\HepParticle{a}{}{}}\xspace}
\providecommand{\Ph}{{\HepParticle{h}{}{}}\xspace}
\newcommand{\PHsm}{\ensuremath{\PH_{125}}\xspace}
\newcommand{\PHo}{{\HepParticle{\PH}{0}{}}\xspace}

\newcommand{\hbb}{\ensuremath{\PH\to\bbbar}\xspace}
\newcommand{\qqqq}{\ensuremath{4\Pq}\xspace}
\newcommand{\ww}{\ensuremath{\PW\PW}\xspace}
\newcommand{\zz}{\ensuremath{\PZ\PZ}\xspace}
\newcommand{\vv}{\ensuremath{\PV\PV}\xspace}
\newcommand{\bbww}{\ensuremath{\bbbar\PW\PW}\xspace}
\newcommand{\bbvv}{\ensuremath{\bbbar\PV\PV}\xspace}
\newcommand{\vvq}{\ensuremath{\PV\PV\to\qqqq}\xspace}
\newcommand{\vqq}{\ensuremath{\PV\to\Pq\PAQq}\xspace}
\newcommand{\yww}{\ensuremath{\PY\to\ww}\xspace}
\newcommand{\yzz}{\ensuremath{\PY\to\zz}\xspace}
\newcommand{\yvv}{\ensuremath{\PY\to\vv}\xspace}
\newcommand{\yvvq}{\ensuremath{\PY\to\vv\to\qqqq}\xspace}
\newcommand{\yfourq}{\ensuremath{\PY\to\qqqq}\xspace}
\newcommand{\XHY}{\ensuremath{{\PX\to\PH\PY}}\xspace}
\newcommand{\XHYbbWW}{\ensuremath{{\PX\to\PH\PY\to\bbww}}\xspace}
\newcommand{\XHYbbVV}{\ensuremath{{\PX\to\PH\PY\to\bbvv}}\xspace}

\newcommand{\TXbb}{\ensuremath{T_\text{Xbb}}\xspace}
\newcommand{\TWqq}{\ensuremath{T_\text{Wqq}}\xspace}
\newcommand{\TXbbbb}{\ensuremath{T_\text{Xbb}^\mathrm{bb}}\xspace}
\newcommand{\TYWW}{\ensuremath{T_\text{YWW}}\xspace}
\newcommand{\TYWWww}{\ensuremath{T_\text{YWW}^\mathrm{VV}}\xspace}
\newcommand{\PQCD}{\ensuremath{P_\text{QCD}}\xspace}
\newcommand{\PTop}{\ensuremath{P_\text{top}}\xspace}
\newcommand{\PHWWqqqq}{\ensuremath{P_\text{HWW4q}}\xspace}
\newcommand{\PHWWqqq}{\ensuremath{P_\text{HWW3q}}\xspace}

\newcommand{\sstat}{\ensuremath{{\sigma_\text{stat}}}\xspace}
\newcommand{\ssyst}{\ensuremath{{\sigma_\text{syst}}}\xspace}

\newcommand{\mx}{\ensuremath{{m_\PX}}\xspace}
\newcommand{\mxrec}{\ensuremath{{m_\PX^\text{rec}}}\xspace}
\newcommand{\mh}{\ensuremath{{m_\PH}}\xspace}
\newcommand{\my}{\ensuremath{{m_\PY}}\xspace}
\newcommand{\myrec}{\ensuremath{{m_\PY^\text{rec}}}\xspace}
\newcommand{\mxmy}{\ensuremath{{\mx,\my}}\xspace}
\newcommand{\mxmyij}{\ensuremath{{m_\PX^i,m_\PY^j}}\xspace}

\newcommand{\mreg}{\ensuremath{{m_\text{reg}}}\xspace}
\newcommand{\mregbb}{\ensuremath{{m_\text{reg}^\mathrm{bb}}}\xspace}
\newcommand{\mregvv}{\ensuremath{{m_\text{reg}^\mathrm{VV}}}\xspace}

\newcommand{\rpf}{\ensuremath{R}\xspace}
\newcommand{\rpfmxmy}{\ensuremath{R(\mxmy)}\xspace}
\newcommand{\rpfij}{\ensuremath{R(\mxmyij)}\xspace}
\newcommand{\rpfdata}{\ensuremath{R^\text{data}(\mxmy)}\xspace}
\newcommand{\rpfmc}{\ensuremath{R^\text{sim}(\mxmy)}\xspace}
\newcommand{\nqcdfail}{\ensuremath{N_\mathrm{QCD\ fail}^{ij}}\xspace}
\newcommand{\nqcdpass}{\ensuremath{N_\mathrm{QCD\ pass}^{ij}}\xspace}
\newcommand{\nmx}{\ensuremath{n_\mx}\xspace}
\newcommand{\nmy}{\ensuremath{n_\my}\xspace}
\newcommand{\bmx}{\ensuremath{b_{k, \nmx}(\mx)}\xspace}
\newcommand{\bmy}{\ensuremath{b_{l, \nmy}(\my)}\xspace}
\newcommand{\akl}{\ensuremath{a_{k,l}}\xspace}

\newcommand{\partmodel}{\textsc{ParT}\xspace}
\newcommand{\particlenet}{\textsc{ParticleNet}\xspace}

\newcommand{\ifrac}[2]{\ensuremath{{#1}/{#2}}\xspace}
\newcommand{\CP}{\ensuremath{CP}\xspace}

\cmsNoteHeader{B2G-23-007}

\title{Search for heavy scalar resonances decaying to Lorentz-boosted Higgs and Higgs-like bosons in the \texorpdfstring{$\bbbar4\Pq$}{bbbar4q} final state at \texorpdfstring{$\sqrt{s} = 13\TeV$}{sqrt(s) = 13 TeV}}

\date{\today}

\abstract{A search is performed for a heavy scalar resonance \PX decaying to a Higgs boson (\PH) and a Higgs-like scalar boson (\PY) in the two bottom quark (\hbb) and four quark (\yvvq) final state, where \PV denotes a \PW or \PZ boson. Masses of the \PX between 900 and 4000\GeV and the \PY between 60 and 2800\GeV are considered. The search is performed in data collected by the CMS experiment at the CERN LHC from proton-proton collisions at 13\TeV center-of-mass energy, with a data set corresponding to a total integrated luminosity of 138\fbinv. It targets the Lorentz-boosted regime, in which the products of the \hbb decay can be reconstructed as a single large-area jet, and those from the \yvvq decay as either one \yfourq or two \vqq jets. Jet identification and mass reconstruction exploit machine-learning tools, including a novel attention-based ``particle transformer'' for \yfourq identification. No significant excess is observed in the data above the standard model background expectation. Upper limits on the product of production cross section and branching fraction as low as 0.2\unit{fb} are derived at 95\% confidence level for various mass hypotheses. This is the first search at the LHC for scalar resonances in the all-hadronic \bbvv decay channel.
}

\hypersetup{
    pdfauthor={CMS Collaboration},
    pdftitle={Search for heavy scalar resonances decaying to Lorentz-boosted Higgs and Higgs-like bosons in the bb4q final state at sqrt(s) = 13TeV},
    pdfsubject={CMS},
    pdfkeywords={CMS, exotica, boosted, diboson, resonance, X, H, Y}
}

\maketitle

\section{Introduction}

 {\tolerance=800
  Many theoretical models predict a richer scalar sector than that of the standard model to address its foundational and observational inconsistencies, such as the Higgs boson (\PH) mass hierarchy problem and the baryon asymmetry in the universe.
  These include two-Higgs-doublet models~\cite{Branco:2011iw} that add a scalar doublet to the standard model, such as the minimal supersymmetric extension of the standard model (MSSM)~\cite{Craig:2013hca}.
  The MSSM predicts two neutral \CP-even scalars (\PHo, \Ph), one neutral \CP-odd scalar (\PA), and two charged scalars ($\PH^\pm$), where one of the neutral \CP-even scalars may be the state observed at 125\GeV.
  The next-to-MSSM~\cite{Domingo:2022kfm} adds to this a complex scalar singlet, predicting additional \CP-even and \CP-odd neutral scalars.
  Finally, the two-real-singlet model~\cite{Robens:2020} predicts two additional \CP-even scalar fields.
  Depending on the masses of the additional scalar particles, all of these models allow the decays of a heavier scalar to symmetric and asymmetric pairs of light scalars, such as $\PH\PH$ and $\Ph\PH$, respectively.
  \par}

Motivated by these theoretical models, we search for decays of the form \XHY, where \PX is the heavier and \PY the lighter new scalar resonance, with \PH decaying to a bottom quark-antiquark pair, \bbbar, and \PY to four quarks via \vvq, where \PV denotes a \PW or \PZ boson.
Many models, such as the two-real-singlet model, predict that the branching fractions of the lighter scalar boson to different standard model particles are similar to or the same as those of the standard model Higgs boson.
Assuming the same mass-dependent branching fractions as the standard model Higgs boson~\cite{LHCHiggsXSWG}, the \ww and \zz decay modes are dominant for masses of the \PY boson $\my > 140\GeV$ and, hence, the \hbb and \yvv decays will be the dominant signatures for the \XHY signal.

\begin{figure}[htpb]
    \centering

    \includegraphics[width=0.95\linewidth,clip=true,viewport=0 200 1920 880]{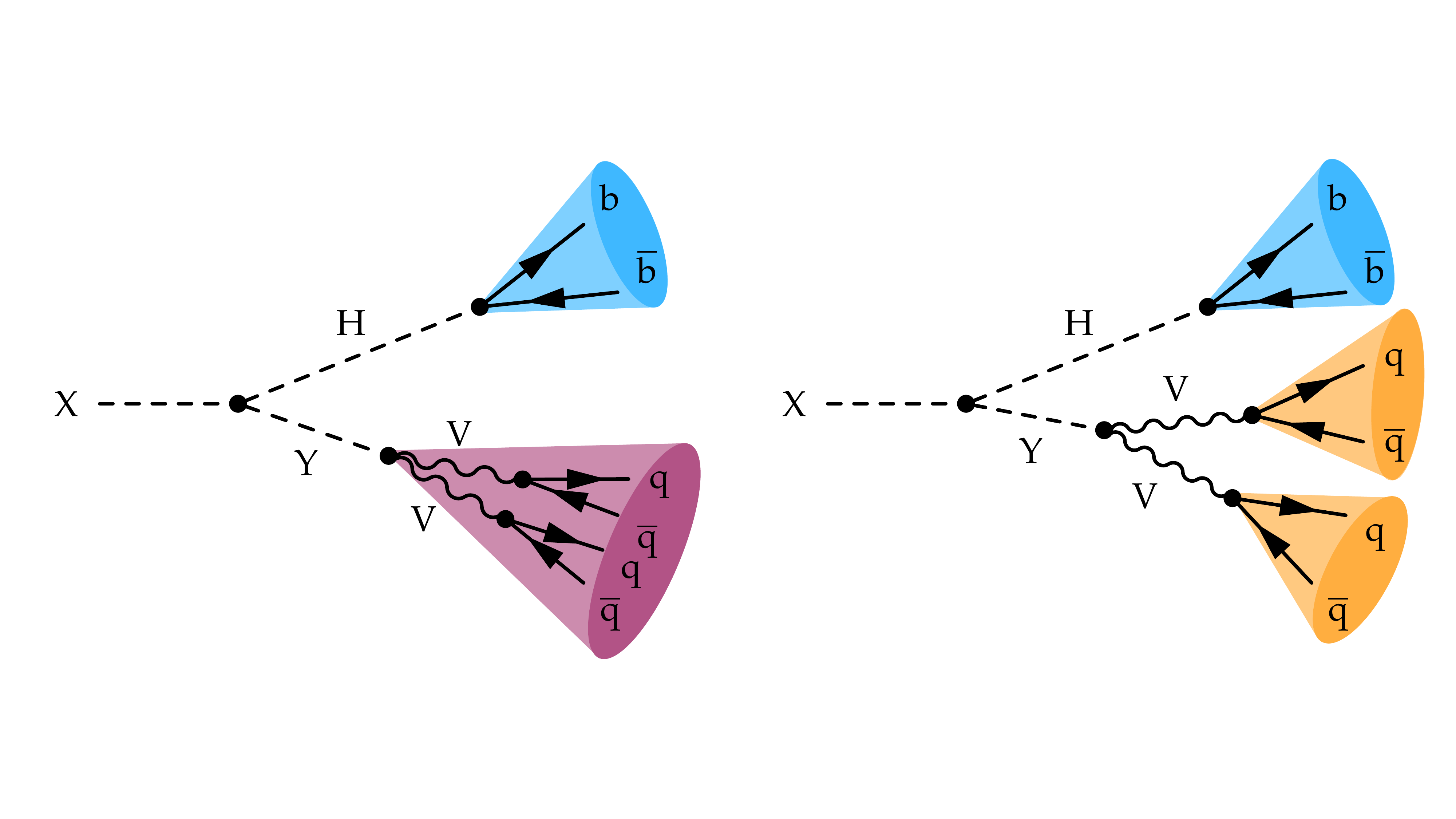}

    \caption{Schematic diagrams of the \XHY decay in the \hbb and \yvvq all-hadronic final state, with $\PV=\PW$ or \PZ.
        Left: The fully merged regime ($\my \lesssim 0.1 \mx$) in which \PY is highly boosted, with its decay products merged and reconstructed as a single large-area jet.
        Right: The semimerged regime ($0.1\mx \lesssim \my < \mx - \mh$) in which \PY is not as boosted, and each daughter \PV boson is instead reconstructed individually as a large-area jet.
    }
    \label{fig:diagrams}
\end{figure}

We target the high \PX  mass (\mx) regime, in which the Higgs boson is highly Lorentz-boosted, with the \hbb system reconstructed as a large-area jet.
Two regimes are considered for the \yvv decay: one where $\my \lesssim 0.1 \mx$, resulting in a highly boosted \PY with its \vvq decay products merged and reconstructed as a single large-area jet (``fully merged''); and the second where $0.1\mx \lesssim \my < \mx - \mh$, in which \PY is not as boosted, and each daughter \PV is instead reconstructed individually as a large-area jet (``semimerged'').
Schematic diagrams of the decays targeted in the two regimes are depicted in Fig.~\ref{fig:diagrams}.

There have been several searches for \XHY production in a variety of kinematic regimes and final states with the data collected by the CMS experiment at the CERN LHC between 2016--2018, such as the merged 4\Pb final state~\cite{CMS:2022suh} and the resolved $\bbbar\PGg\PGg$~\cite{CMS:2023boe} and $\bbbar\PGt\PGt$~\cite{CMS:2021yci} final states.
Similarly, ATLAS has searched for this process in the $\PGg\PGg\bbbar$~\cite{atlas-bbgg} final state, along with the $\PGg\PGg\PV\PV$~\cite{atlas-ggVV} and $\PGt\PGt\PV\PV$~\cite{atlas-tautauVV} final states.
Previous searches at CMS in the \bbvv final state have only been conducted in symmetric decays of the form $\PX \to \PH\PH$~\cite{CMS:2024rgy}, and only in the semileptonic channel.

In this paper, we present the first search for asymmetric cascade decays of the form \XHY in the all-hadronic \bbvv final state.
The analysis is performed using data collected by CMS from proton-proton ($\Pp\Pp$) collisions at $\sqrt{s}=13\TeV$ in 2016--2018, with a data set corresponding to a total integrated luminosity of 138\fbinv~\cite{CMS-LUM-17-003,CMS-PAS-LUM-20-001}.

The paper is organized as follows.
Section~\ref{sec:detector} provides a brief overview of the CMS detector.
Section~\ref{sec:EventSimulation} describes the simulated samples used in this analysis, followed by a description of the event reconstruction and techniques used to improve the jet mass resolution in Section~\ref{sec:EventReconstruction}.
The event selection and object identification criteria for the fully merged and semimerged categories are outlined in Section~\ref{sec:EventSelection}.
A description of the background estimation method and its validation is provided in Section~\ref{sec:BackgroundEstimation}.
Section~\ref{sec:SystUnc} summarizes the sources of systematic uncertainty in the analysis.
Finally, the results are presented in Section~\ref{sec:Results} and summarized in Section~\ref{sec:Summary}.
Tabulated results are presented in the HEPData record for this analysis~\cite{hepdata}.

\section{The CMS detector}
\label{sec:detector}

The central feature of the CMS apparatus is a superconducting solenoid of 6\unit{m} internal diameter, providing a magnetic field of 3.8\unit{T}.
Within the solenoid volume are a silicon pixel and strip tracker, a lead tungstate crystal electromagnetic calorimeter (ECAL), and a brass and scintillator hadron calorimeter (HCAL), each composed of a barrel and two endcap sections.
Forward calorimeters extend the pseudorapidity ($\eta$) coverage provided by the barrel and endcap detectors.
Muons are reconstructed using gas-ionization detectors interleaved with the layers of the steel flux-return yoke outside the solenoid.
More detailed descriptions of the CMS detector, together with a definition of the coordinate system used and the relevant kinematic variables, can be found in Refs.~\cite{CMS:2008xjf, CMS:2023gfb}.

Events of interest are selected using a two-tiered trigger system.
The first level, composed of custom hardware processors, uses information from the calorimeters and muon detectors to select events at a rate of around 100\unit{kHz} within a fixed latency of 4\mus~\cite{CMS:2020cmk}.
The second level, known as the high-level trigger, consists of a farm of processors running a version of the full event reconstruction software optimized for fast processing, and reduces the event rate to a few kHz before data storage~\cite{CMS:2016ngn,CMS:2024aqx}.

\section{Event simulation}
\label{sec:EventSimulation}

The \XHY signal samples are simulated via gluon fusion at leading order in the 900--4000\GeV \mx range and 60--2800\GeV \my range, with widths of 1\MeV each, using the \MGvATNLO \textsc{2.6.5}~\cite{Alwall:2014hca} event generator.
The \hbb and $\PY\to\ww\to\qqqq$ decays are simulated with \PYTHIA ({v8.205})~\cite{Sjostrand:2014zea}.
Signal samples are produced in the next-to-MSSM with the narrow-width approximation, which allows for the interpretation of the results in the context of other scenarios beyond the standard model under the same approximation.
The simulation of parton showering, underlying event interactions, and hadronization is performed with \PYTHIA ({v8.205}), including parameters from the CP5 tune~\cite{CMS:2019csb}.

The primary background is composed of jets produced through the strong interaction, referred to as quantum chromodynamics (QCD) multijet events.
It is simulated with two to four jets in the final state and at leading order with \MGvATNLO \textsc{2.2.2}, though it is estimated primarily via control samples in data.
Minor backgrounds include top quark (\PQt) events, such as {\ttbar}+jets and single {\PQt}+jets, which are simulated with \textsc{Powheg}\,2.0~\cite{POWHEG1, Frixione:2007nw, POWHEG2, POWHEG3, POWHEG-TT} and \PYTHIA\textsc{8}; {\PV}+jets and diboson events, simulated with \MGvATNLO and \PYTHIA\textsc{8}; and standard model \PH events, dominated by associated production with a \PV boson and simulated with \textsc{Powheg}\,2.0 and \PYTHIAviii.
All simulated samples are generated with the NNPDF 3.0~\cite{NNPDF:2014otw} or 3.1~\cite{NNPDF:2017mvq} parton distribution functions (PDFs).
The response of the CMS detector is modeled with the \GEANTfour~\cite{GEANT4:2002zbu} package, with additional simulated $\Pp\Pp$ interactions superimposed on the hard interactions in simulated events to model pileup effects.
The distribution of the number of pileup interactions in the simulation is reweighted to match the corresponding distribution in data, which is obtained using the measured instantaneous luminosity per bunch crossing.

\section{Event reconstruction}
\label{sec:EventReconstruction}

A particle-flow algorithm~\cite{CMS:2017yfk} aims to reconstruct and identify each individual particle in an event, with an optimized combination of information from the various elements of the CMS detector.
The energy of photons is obtained from the ECAL measurement.
The energy of electrons is determined from a combination of the electron momentum at the primary interaction vertex as determined by the tracker, the energy of the corresponding ECAL cluster, and the energy sum of all bremsstrahlung photons spatially compatible with originating from the electron track.
The energy of muons is obtained from the curvature of the corresponding track.
The energy of charged hadrons is determined from a combination of their momentum measured in the tracker and the matching ECAL and HCAL energy deposits, corrected for the response function of the calorimeters to hadronic showers.
Finally, the energy of neutral hadrons is obtained from the corresponding corrected ECAL and HCAL energies.
The primary vertex is taken to be the vertex corresponding to the hardest scattering in the event, evaluated using tracking information alone, as described in Section 9.4.1 of Ref.~\cite{CMS-TDR-15-02}.

For each event, hadronic jets are clustered from these reconstructed particles using the infrared- and collinear-safe anti-\kt algorithm~\cite{Cacciari:2008gp, Cacciari:2011ma} with a distance parameter of 0.8 (AK8 jets).
Jet momentum is determined as the vectorial sum of all particle momenta in the jet, and is found from simulation to be, on average, within 5 to 10\% of the generator jet momentum over the entire transverse momentum (\pt) spectrum and detector acceptance~\cite{CMS:2016lmd}.
Additional $\Pp\Pp$ interactions within the same or nearby bunch crossings (pileup) can contribute additional tracks and calorimetric energy depositions to the jet momentum.
The pileup-per-particle identification algorithm~\cite{Bertolini:2014bba} is used to mitigate the effect of pileup at the reconstructed particle level, making use of local shape information, event pileup properties, and tracking information.
Additional selection criteria are applied to each jet to remove those potentially dominated by instrumental effects or reconstruction failures. A comprehensive description of these criteria and the full jet momentum calibration procedure is provided in Ref.~\cite{Sirunyan:2020foa}.

Jet mass is reconstructed by different algorithms, depending on the type of jet and context.
In the high-level trigger, the ``trimmed mass'' is defined as the invariant mass of the jet constituents after a jet trimming procedure~\cite{Krohn:2009th} to remove soft and wide-angle radiation, using a subjet size parameter of 0.3 and a subjet to AK8 jet \pt fraction of 0.1.
Offline, a mass regression algorithm is employed, which uses the \particlenet architecture~\cite{Qu:2019gqs, CMS-BTV-22-001}, leveraging particle-flow candidate and secondary vertex information in a graph neural network to predict the ``regressed jet mass'' (\mreg)~\cite{CMS-DP-2021-017}.
The \particlenet model has demonstrated improved mass resolution for AK8 jets by up to 10\% compared to alternative algorithms~\cite{CMS-DP-2021-017}.

\section{Event selection}
\label{sec:EventSelection}

The online selection consists of triggers requiring events with a scalar sum of jet transverse momenta, \HT, of at least 800, 900, or 1050\GeV, depending on the data-taking period.
Further trigger selection records events with at least one AK8 jet with a $\pt \geq 400$ or 500\GeV, where in the case of the former, an additional requirement that the jet has a trimmed mass of at least 40\GeV is imposed.
Finally, triggers requiring an \HT of 750 or 800\GeV with all jets having trimmed masses of at least 50\GeV are also used to target the boosted, fully hadronic nature of the final state.
The efficiency for the combined set of triggers is measured in data in a sample composed primarily of AK8 jets originating from hadronically decaying top quarks, as a function of jet \pt and mass, and is applied to simulated background and signal processes.
The signal efficiency of the online selection ranges from 90\% for low $\mx \approx 1000\GeV$ to close to 100\% for higher $\mx \gtrsim 1500\GeV$.

The \hbb decay is targeted through an offline selection for a single AK8 jet, while the \yvv decay is selected as either one or two AK8 jets, in the fully merged and semimerged categories, respectively.
In both cases, \particlenet is used to discriminate signal \hbb jets from QCD background using a discriminant \TXbb derived from its outputs.
These outputs have been decorrelated from the jet masses by enforcing a uniform distribution in jet mass and \pt in the training samples~\cite{CMS-DP-2020-002}.
Three \TXbb working points (WPs, adjusted for each data-taking year~\cite{CMS-BTV-22-001}) are considered in the analysis: ``low'', ``medium'', and ``high'' purity (LP, MP, and HP), corresponding to signal (QCD background) efficiencies of approximately 80 (0.8), 60 (0.3), and 40 (0.1)\%, respectively. An overview of the training, mass decorrelation, and calibration of the \particlenet \hbb tagger can be found in Ref.~\cite{CMS-BTV-22-001}.
As jet mass resolution is crucial to the sensitivity of the search, the regressed mass observable \mreg is used for all AK8 jets.

The \PV bosons in the simulated \yvv samples, used to train the jet classification algorithms for both the fully merged and semimerged categories of events, are \PW bosons that decay hadronically.
As a result, the sensitivity to the \yzz decay mode is less than that for \yww.
The selection efficiency for signals with the \yzz decay is measured in a subset of signal samples to be a factor of 0.73 and 0.78 times that of \ww decays in the fully merged and semimerged categories, respectively.

\begin{figure}[htbp!]
    \centering

    \includegraphics[width=0.495\textwidth]{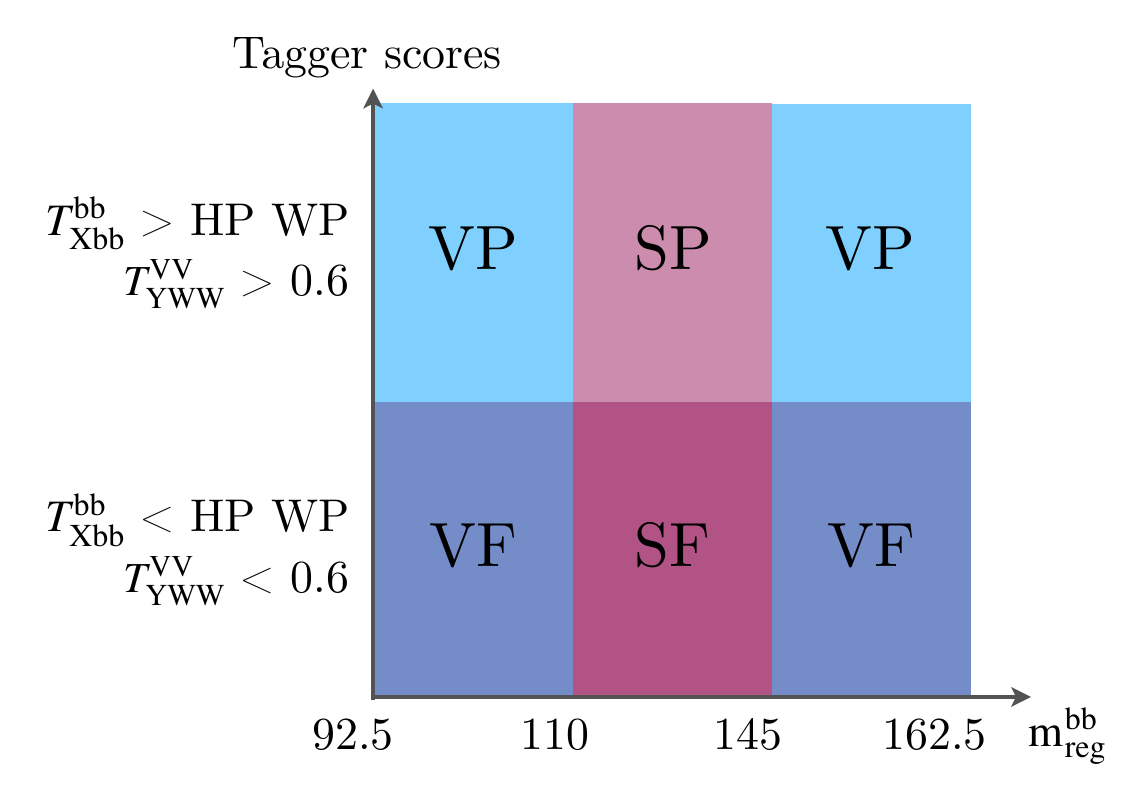}
    \includegraphics[width=0.495\textwidth]{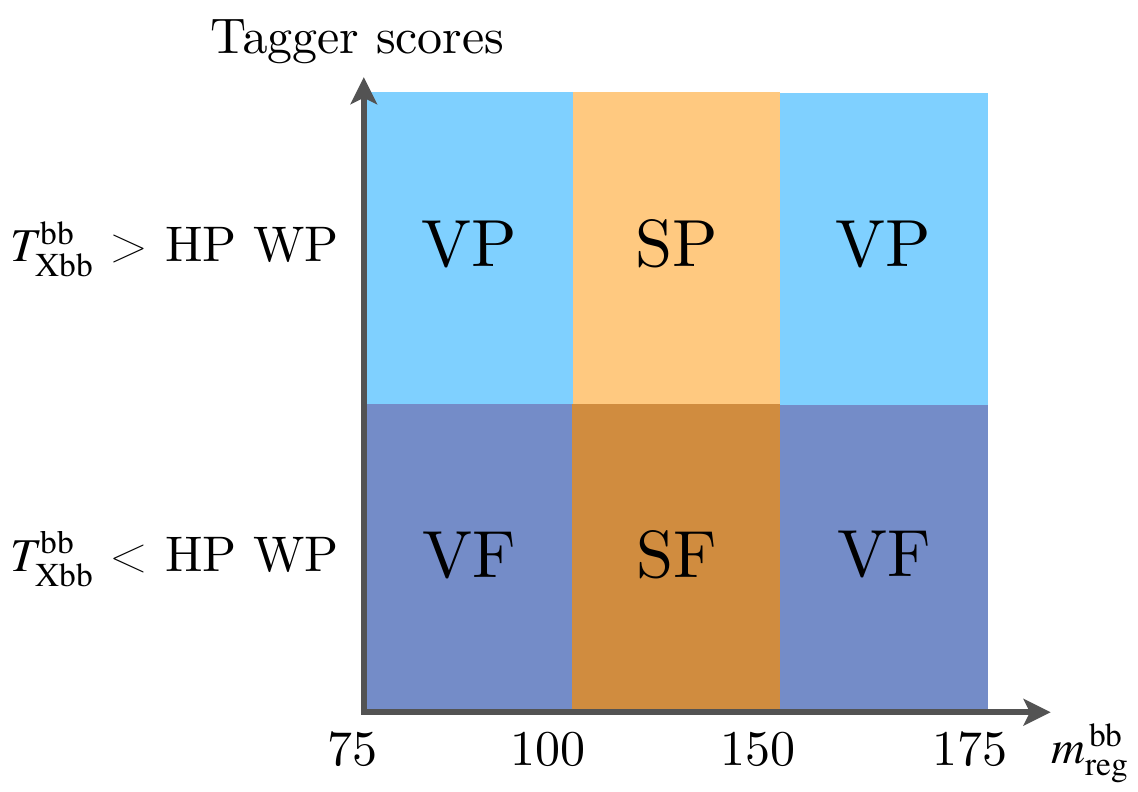}

    \caption{Schematic of the event categorization into the signal pass (SP), signal fail (SF), validation pass (VP), and validation fail (VF) regions for the fully merged (left) and semimerged (right) categories. The pass and fail tagging regions of the fully merged and semimerged categories are determined by application of the high purity (HP) working point (WP) of the \TXbb tagger to the \bbbar candidate jet (\TXbbbb). In the fully merged category, the pass and fail tagging regions are additionally defined by application of the \TYWW tagger to the \yvvq candidate jet (\TYWWww).}
    \label{fig:EventCategorization}
\end{figure}

For both categories, a ``signal pass`` (SP) region is optimized for performing the search.
An independent ``signal fail'' (SF) region with the tagger selections inverted is also used as a control region to estimate the QCD multijet background, as described in Section~\ref{sec:BackgroundEstimation}.
Along with these signal regions (SRs), which require the \bbbar candidate \mreg (\mregbb) to be within a window around \mh, we define validation pass (VP) and validation fail (VF) regions by the sidebands of this mass window to validate the background estimation method before observing data in the SRs.
These validation regions (VRs) are used purely to confirm the agreement of the background model with data and are not used in the final search.
Signal contamination in the VRs and the SF region is ${<}1\%$ for both the fully merged and semimerged categories, assuming signal cross sections corresponding approximately to those obtained from scans of the next-to-MSSM benchmark planes across the range of \mx and \my considered in Ref.~\cite{Ellwanger:2022jtd}.
Detailed selections for all four regions for both categories are defined below with a schematic of the event categorization provided in Fig.~\ref{fig:EventCategorization}.

\subsection{Fully merged category}
\label{sec:EventSelectionMergedY}

For $\my \lesssim 0.1\mx$, the fraction of signal \yvv AK8 jets containing four quarks is approximately 80--90\%, and decreases to around 40\% in the region $0.1\mx \lesssim \my \lesssim 0.16\mx$ before rapidly dropping off. Thus, in the case where $\my \lesssim 0.1\mx$, both \PV bosons and at least three out of four of their combined quark decay products are assumed to be reconstructed as a single AK8 jet.
Hence, selected events are required to have at least two AK8 jets with $\abs{\eta} < 2.4$, corresponding to the Higgs and \PY boson candidates.
The leading and subleading (in \pt) jets are required to have $\pt>400$ and 350\GeV, respectively.
Out of all AK8 jets in the event passing these requirements, the one with the highest \TXbb score is then considered as the \hbb candidate jet, and is required to pass the HP WP of the tagger.
This \bbbar candidate jet is also required to have a jet mass close to \mh: $110 < \mreg < 145\gev$.

To identify the \yvvq jet, we employ an attention-based neural network, referred to as ``particle transformer'' (\partmodel)~\cite{Qu:2022mxj}, which is trained on jets originating from scalar \PH-like resonances decaying to \ww, with at least one \PW boson decaying hadronically~\cite{CMS-PAS-JME-25-001}.

The \partmodel algorithm is decorrelated with the jet mass by enforcing a uniform distribution in jet mass, and by varying the masses of the daughter \PV bosons such that the ratio of \PV and \PY boson masses at each \my point is equal to the ratio of \PV and Higgs boson masses in the standard model~\cite{CMS-PAS-JME-25-001}.
Because the mass of the scalar resonance in the training samples only varies between 15--250\GeV, the selection efficiency for signal jets with \my exceeding this range is lower by a factor of three compared to those with \my within this range. A detailed description of the training, mass decorrelation, and performance of this network is provided in Ref.~\cite{CMS-PAS-JME-25-001}.

The outputs of \partmodel are used to derive the \TYWW discriminant, isolating the \yvv signal from QCD and \PQt background jets, defined as:
\begin{equation}
    \label{eq:thww}
    \TYWW = \frac{\PHWWqqqq + \PHWWqqq}{\PQCD + \PTop + \PHWWqqqq + \PHWWqqq},
\end{equation}
where \PHWWqqqq, \PHWWqqq, \PQCD, and \PTop are the predicted probabilities of the jet being a four-pronged \vv-jet, three-pronged \vv-jet, QCD jet, or \PQt jet, respectively.
The AK8 jet with the highest \TYWW score that also passes the above \pt and $\eta$ kinematic selections is considered the \vv candidate jet.
It is required to have $\TYWW > 0.6$, corresponding to an approximately 60 (1)\% signal (background) efficiency for jets with $\my < 250\GeV$, which is the boundary of the training mass range. The performance of the tagger at this working point is similar for all \my beneath this threshold, as shown in Ref.~\cite{CMS-PAS-JME-25-001}.
If the same AK8 jet has both the highest \TXbb and \TYWW score in the event, it is classified as the \bbbar candidate and the next-highest \TYWW scoring jet is considered the \vv candidate, where the \TXbb score  of the \bbbar candidate and the \TYWW score of the \vv candidate are denoted \TXbbbb and \TYWWww, respectively.
The fraction of signal events in which this occurs and that pass our offline selections is found to be less than 1\%.
All the \pt and tagger selections are jointly optimized for the most stringent expected exclusion limits for a range of \mx and \my points.

The search is performed in events passing these selections, referred to as the SP region, binned in the two-dimensional (2D) plane of the \vv candidate jet's regressed mass (\mregvv) and the invariant mass of the system composed of the \bbbar and \vv candidate jets, representing the reconstructed \my and \mx (\myrec and \mxrec), respectively.
The SF region, used to estimate the QCD multijet background, is defined by inverting the tagger selections for both jets.
Finally, separate VP and VF regions, defined using the \mregbb sidebands, are used to validate the background estimation technique.

\subsection{Semimerged category}
\label{sec:EventSelectionResolvedY}

The semimerged search targets the $0.1\mx \lesssim \my < \mx$ mass range where the hadronic decays of the two \PV bosons produce two distinct AK8 jets, as opposed to a single jet like in the fully merged category.
Therefore, the semimerged category selects events containing at least three AK8 jets---each with $\pt > 200 \GeV$, $\abs{\eta} < 2.4$, and $\mreg > 50\GeV$---including an AK8 jet from the \hbb decay and two AK8 jets from the \PV bosons.
The \particlenet model is used to target the \hbb jet, as well as the two \vqq jets from the \PY decay using a discriminant \TWqq derived from its outputs.

Of the three highest \pt AK8 jets that pass the above selection, the jet with the highest \particlenet \TXbb score is considered the \bbbar candidate jet, and is required to have $100 < \mreg < 150\GeV$ ($75 < \mreg < 100\GeV$ or $150 < \mreg < 175\GeV$) in the SR (VR).
The two remaining AK8 jets are considered \vqq candidates, and must have $70 < \mreg < 100\GeV$ and pass a loose working point of the \TWqq discriminant yielding a signal efficiency that ranges from 60 to 95\% for jet \pt between 200 and 800\GeV.

The SP region of the semimerged category is then defined by requiring that the \TXbb score of the \bbbar candidate jet (\TXbbbb) be greater than the HP WP.
As in fully merged category, events are binned in the 2D \myrec and \mxrec plane, defined as the invariant masses of the systems of the two \vqq candidate jets and of all three AK8 jets, respectively.
The SF region is formed by inverting the \TXbbbb score requirement, while the VRs are defined using the \mregbb sidebands.
In order to ensure orthogonality of the two categories, events passing the fully merged selections are additionally rejected if they contain two AK8 jets with masses between 60--110\GeV that pass working points of the \TWqq discriminant corresponding to a signal efficiency of 60\%.

\section{Background estimation}
\label{sec:BackgroundEstimation}

The search is performed in the SP regions of the two categories using a 2D binned likelihood model in \mxrec and \myrec.
The background in these regions is dominated by QCD multijet events, whose yield is estimated using the product of data in the fail regions and polynomial ``transfer functions''~\cite{Sirunyan:2017isc,Sirunyan:2018qca,Sirunyan:2020hwz}, as described below.
Minor backgrounds include \PQt and \PV backgrounds, which are modeled using the Monte Carlo (MC) simulations described in Section~\ref{sec:EventSimulation}.
The corresponding systematic and statistical uncertainties are incorporated in the final statistical analysis through nuisance parameters in the likelihood model.

\subsection{Fully merged category}

In the fully merged category, both the \TXbbbb and \TYWWww selections are inverted to increase the sample size in the fail region.
Although \partmodel has also been trained to be decorrelated with the jet mass, there remains a high likelihood that the \TYWWww selection may affect the shape of the \myrec, \ie, \mregvv, observable.
To model the expected effect, we first determine a 2D fail-to-pass transfer function from QCD simulation alone, \rpfmc, which is used in conjunction with a second \rpfdata transfer function to correct for any residual differences in data.
The product \rpfmxmy is the overall fail-to-pass transfer function:
\begin{equation}
    \rpfmxmy = \rpfmc \, \rpfdata,
\end{equation}
where \rpfmc and \rpfdata are both chosen to be 2D Bernstein polynomials in \mx and \my~\cite{bernstein1912}:
\begin{equation}
    \label{eq:bernstein}
    \rpfmxmy = \sum^{\nmx}_{k=0} \sum^{\nmy}_{l=0} \akl \big[ \bmx \, \bmy\big],
\end{equation}
where \nmx and \nmy are the orders of the polynomial in \mx and \my, respectively; and \bmx and \bmy are the $k^\mathrm{th}$ and $l^\mathrm{th}$ Bernstein basis functions of order \nmx and \nmy, respectively.
The coefficients of the polynomial, \akl, are determined through a simultaneous maximum likelihood fit to the data in the fail and pass regions.

\begin{figure}[ht]
    \centering

    \includegraphics[width=0.495\linewidth]{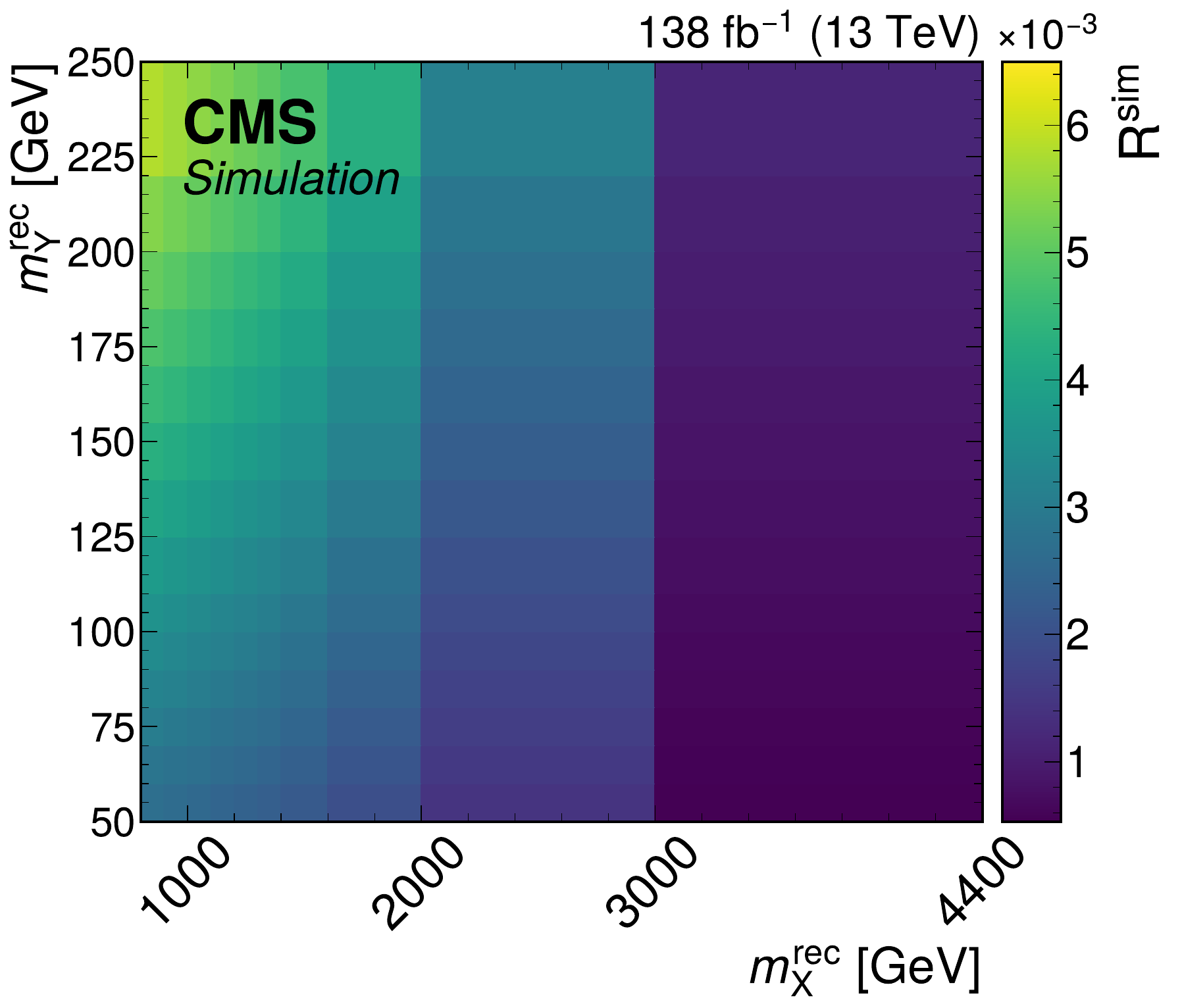}
    \includegraphics[width=0.495\linewidth]{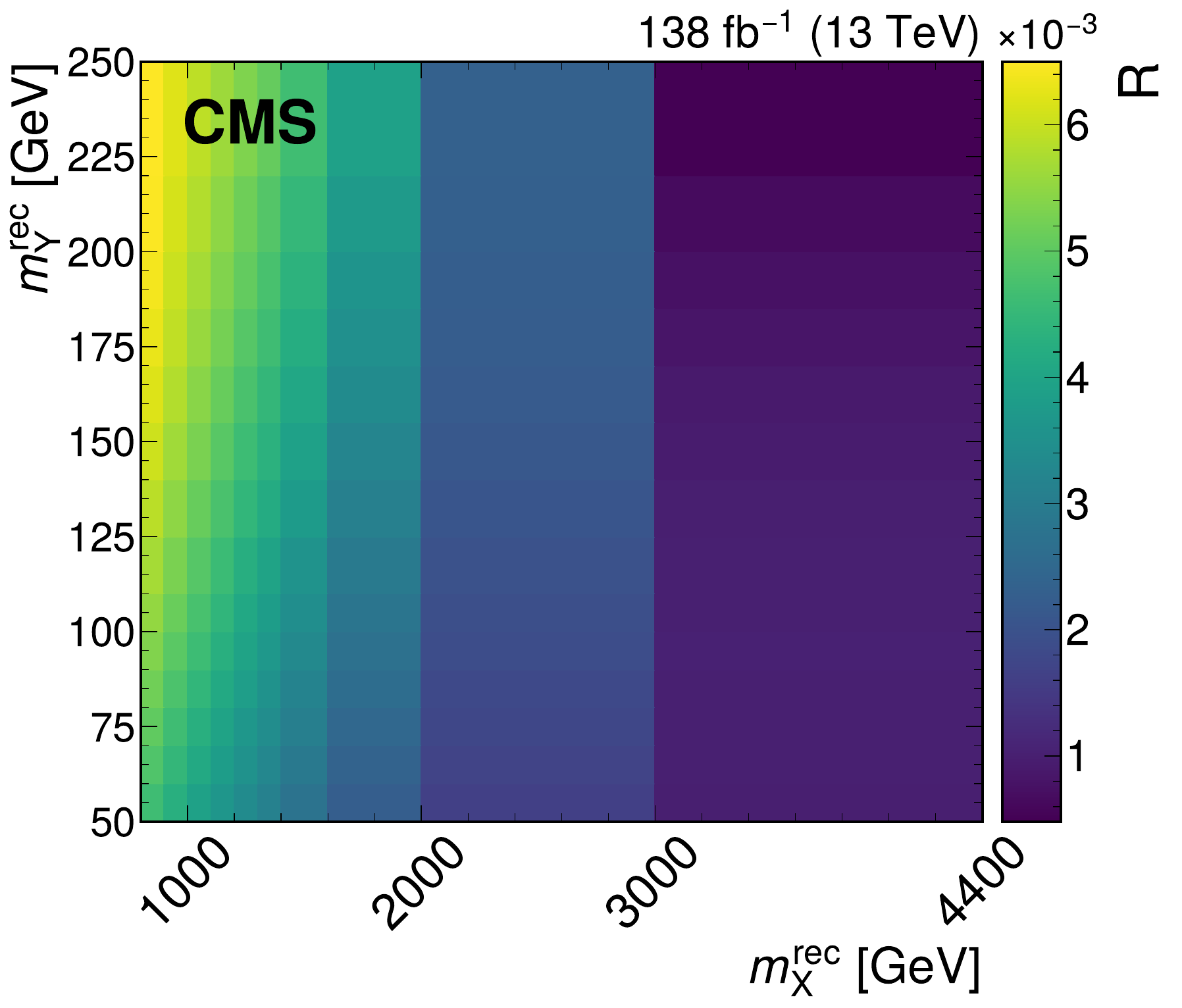}

    \caption{Measured transfer functions \rpfmc (left) and \rpfmxmy (right) between the signal pass (SP) and signal fail (SF) regions of the fully merged category used to estimate the QCD multijet background contribution.}
    \label{fig:BackgroundEstimation_tfs}
\end{figure}

The \rpfmc function is obtained by a dedicated maximum likelihood fit to QCD simulations in the fail and pass regions, with its order in both \mxrec and \myrec determined to be 1 by Fisher F-tests~\cite{FisherFTest} for background-only fits to the data in the signal region.
The \rpfdata is determined through a fit to data in the signal fail and pass regions using the predetermined \rpfmc; its orders are determined to be 2 and 1 in \mxrec and \myrec, respectively.
The measured \rpfmc and total \rpfmxmy transfer functions, as measured in the signal region, are shown in Fig.~\ref{fig:BackgroundEstimation_tfs}, with the same binning in $(\mxmy)$ as used in the maximum likelihood fit to data.
To achieve an adequate sample size in each bin, the bin widths along the $\mx$ axis are gradually increased at higher masses.

\subsection{Semimerged category}

To estimate the QCD multijet background in the pass regions, the data in the respective fail regions are used as independent control regions to model its shape.
In the semimerged category, the SF region is defined by inverting the \TXbbbb selection alone such that the shape of the multijet background regions is obtained directly from the shape in the fail region.

Explicitly, the QCD multijet yield in the $i^\mathrm{th}$ \mxrec and $j^\mathrm{th}$ \myrec bin in the fail regions (binned identically to the pass region) is represented by a floating parameter in the likelihood, \nqcdfail.
The initial value of \nqcdfail is set as the difference between the data and the total simulated backgrounds in that bin.
The QCD yield estimate in the corresponding bin of the pass region, \nqcdpass, is then determined by the product of \nqcdfail and a 2D pass-to-fail transfer function, \rpfmxmy, evaluated at that bin:
\begin{equation}
    \nqcdpass = \rpfij \, \nqcdfail.
\end{equation}
The function \rpf serves to model any potential differences in the shape of the multijet background in the fail and pass regions, with its form chosen to be a 2D Bernstein polynomial in \mx and \my as in Eq.~\ref{eq:bernstein}.
Its order in both \mxrec and \myrec is determined by Fisher F-tests to be 0, \ie, a constant \rpf.
This indicates that the shapes of the QCD multijet background in both regions are indeed consistent, which is not unexpected since the \particlenet tagger has been shown to be successfully decorrelated with the jet mass~\cite{CMS-DP-2020-002}.
The fitted value is $\rpfmxmy = 0.024 \pm 0.002$.

\subsection{Validation}

\begin{figure}[hbtp]
    \centering

    \includegraphics[width=0.48\textwidth]{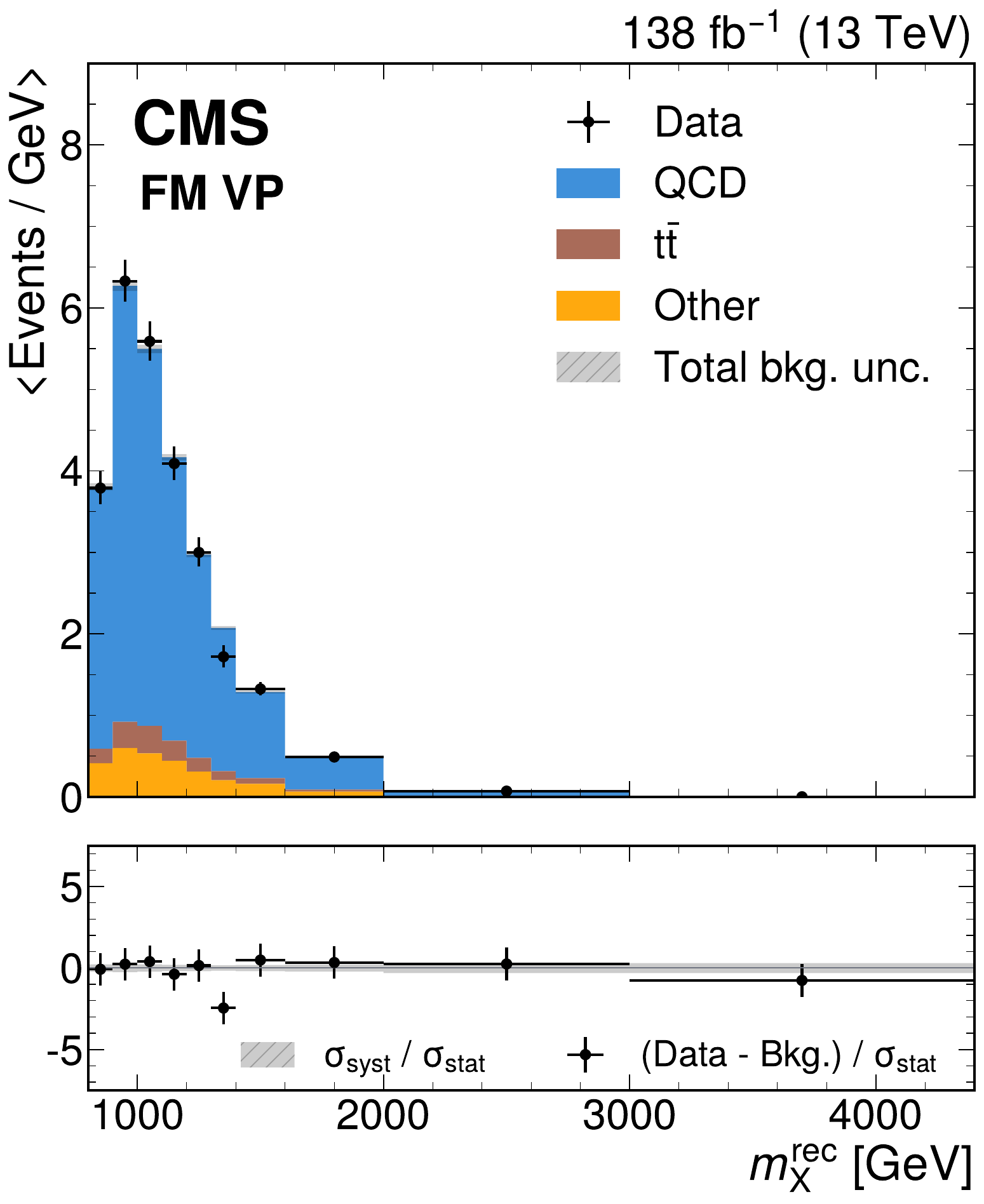}
    \includegraphics[width=0.51\textwidth]{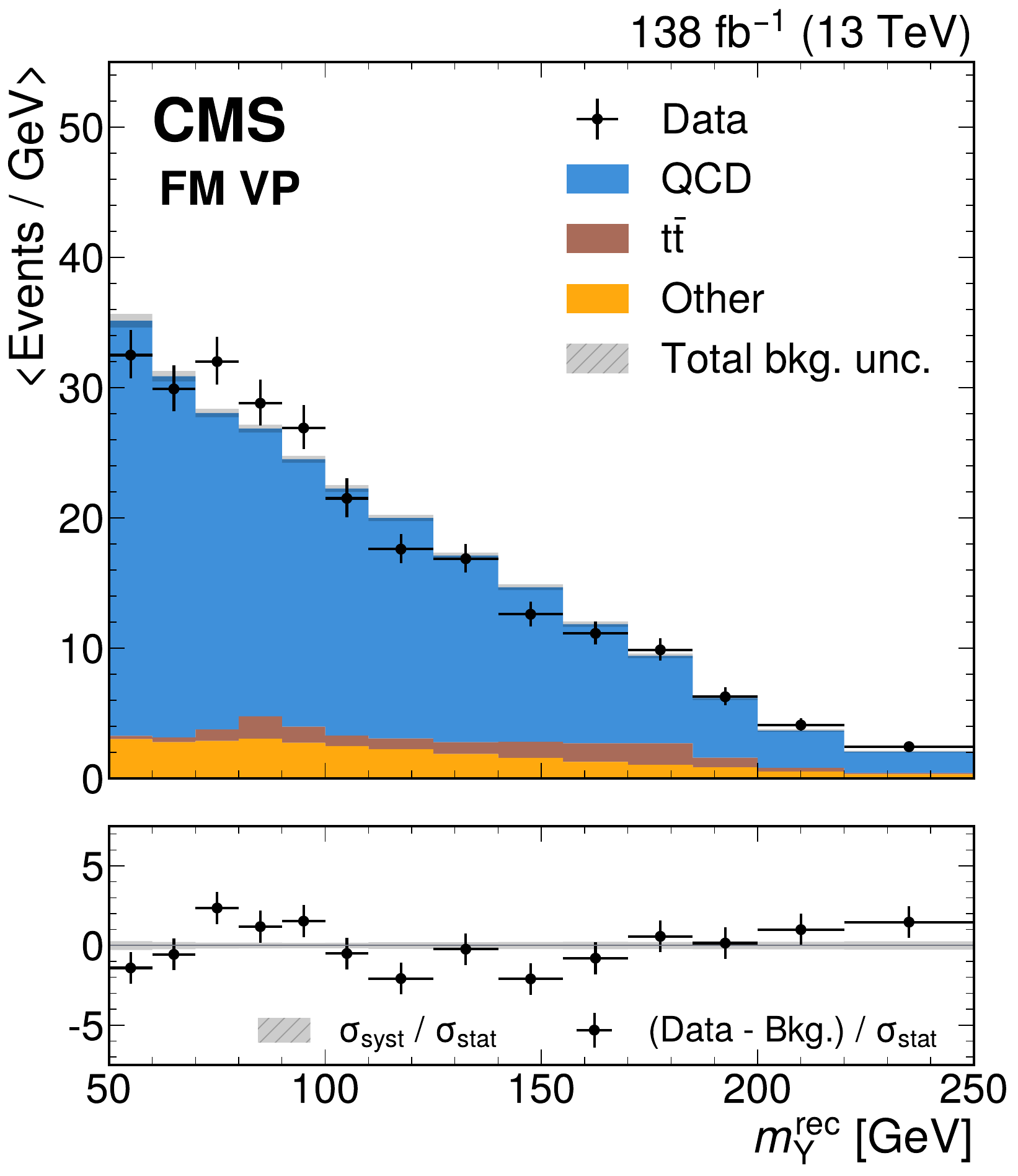}

    \caption{Projected distributions of the \mxrec (left) and \myrec (right) observables in the validation pass (VP) region of the fully merged (FM) category, after a background-only maximum likelihood fit to the data.
        The upper panels show the data and fitted background estimates, as well as the $\pm$1 standard deviation uncertainty in the total background estimate (\ssyst) per bin, divided by the respective bin widths, in gray.
        The vertical error bars for the data represent the statistical uncertainty, while the horizontal bars represent the bin widths.
        The lower panels show the pull per bin, defined as $\ifrac{(\text{data} - \mathrm{Bkg.})}{\sstat}$, as well as a gray band indicating the systematic uncertainty in the pulls, estimated by dividing the corresponding systematic uncertainty in the main panel by $\sstat$.
        \label{fig:bgestimation_postfit_vr_fm}
    }
\end{figure}

\begin{figure}[hbtp]
    \centering

    \includegraphics[width=0.49\textwidth]{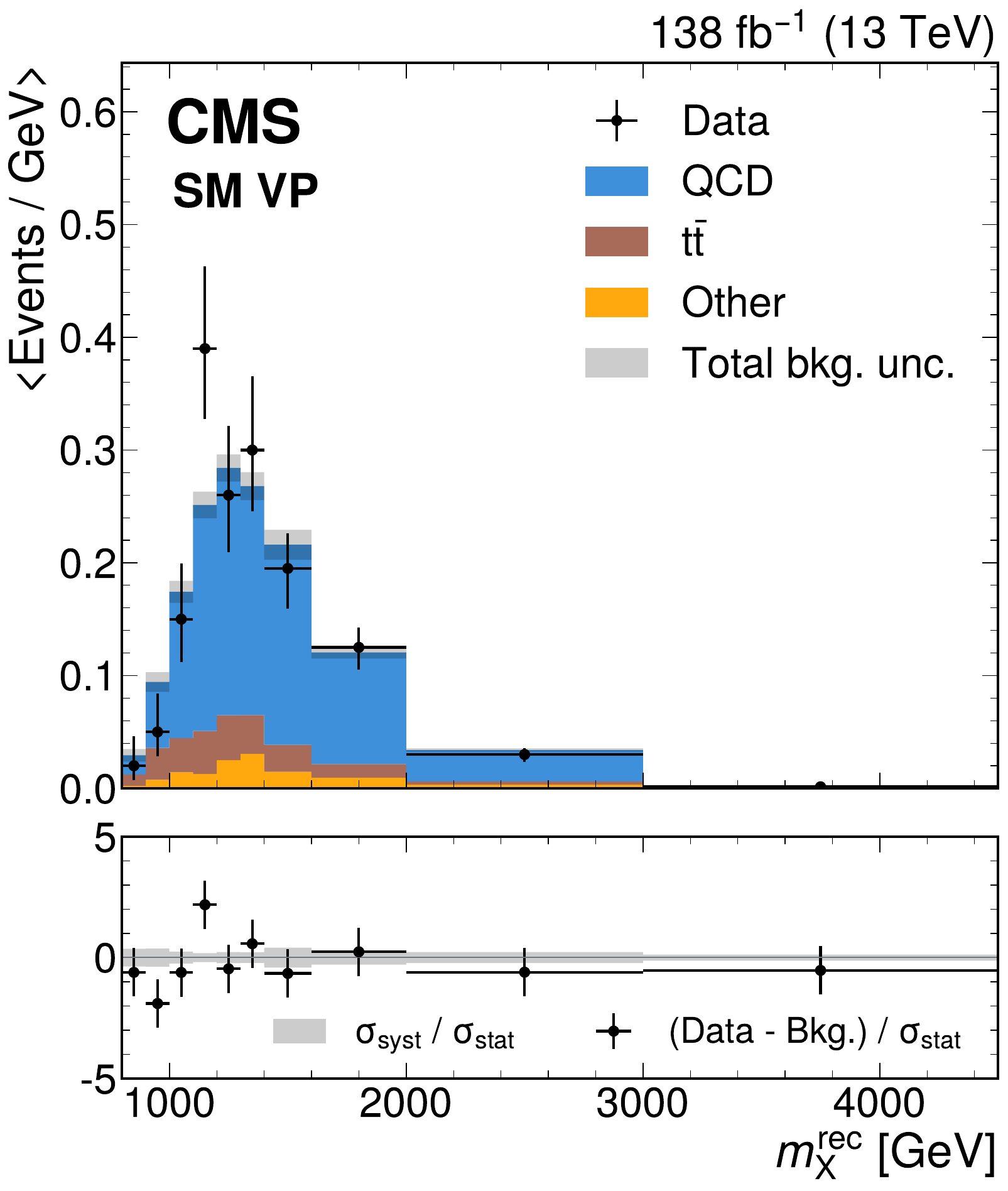}
    \includegraphics[width=0.49\textwidth]{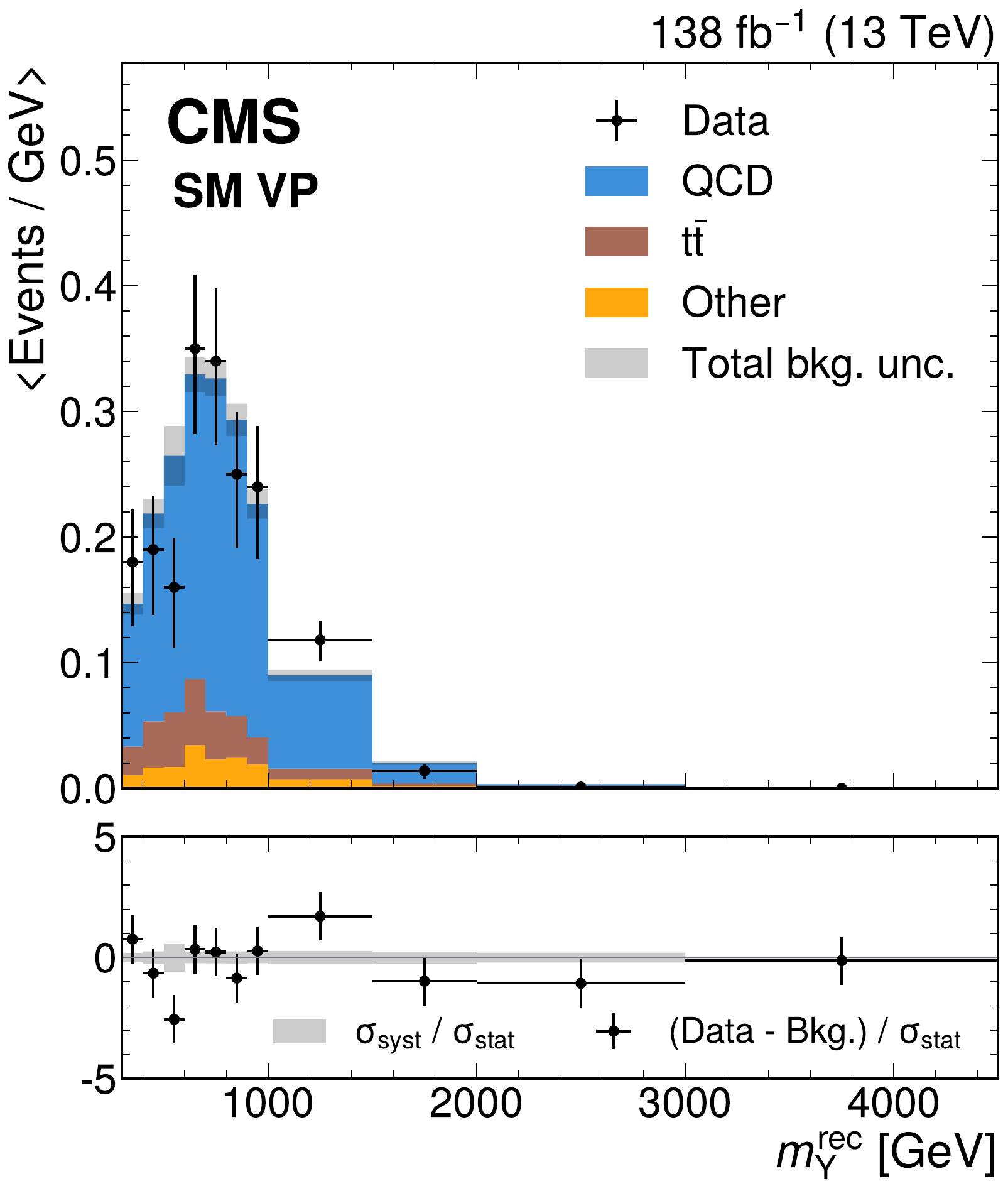}

    \caption{Projected distributions of the \mxrec (left) and \myrec (right) observables in the validation pass (VP) region of the semimerged (SM) category, after a background-only maximum likelihood fit to the data.
        The upper panels show the data and fitted background estimates, as well as the $\pm$1 standard deviation uncertainty in the total background estimate (\ssyst) per bin, divided by the respective bin widths, in gray.
        The vertical error bars for the data represent the statistical uncertainty, while the horizontal bars represent the bin widths.
        The lower panels show the pull per bin, defined as $\ifrac{(\text{data} - \mathrm{Bkg.})}{\sstat}$, as well as a gray band indicating the systematic uncertainty in the pulls, estimated by dividing the corresponding systematic uncertainty in the main panel by $\sstat$.
        \label{fig:bgestimation_postfit_vr_sr}
    }
\end{figure}

{\tolerance=800
    This background estimation procedure is first validated for both categories in dedicated, background-dominated VRs, defined in Section~\ref{sec:EventSelection} and shown in Figs.~\ref{fig:bgestimation_postfit_vr_fm} and~\ref{fig:bgestimation_postfit_vr_sr}.
    The lower panels show the pull per bin, defined as the difference between the data and predicted event yields divided by the statistical uncertainty associated with the data assuming a Poisson distribution, $\sstat$.
    Also shown in the lower panels is a gray band indicating the systematic uncertainty in the pulls, estimated by dividing the corresponding systematic uncertainty in the main panel by $\sstat$.
    The agreement between the data and background model in the VRs is quantified via two-sample goodness of fit tests using the saturated model test statistic~\cite{Baker:1983tu}.
    This yields $p$-values~\cite{Cowan:2010js} of $0.85$ and $0.11$ for the semimerged and fully merged categories, respectively, indicating no significant discrepancies.
    The goodness of fit tests are then repeated for the data and the background models in the SRs, yielding $p$-values of $0.76$ and $0.27$, respectively.
    In both categories, bias tests were performed in pseudodata by injecting various signal mass hypotheses at cross sections corresponding to their respective expected limits. The injected signal cross sections were successfully recovered in all such tests, indicating no significant bias in the signal extraction process.
    \par}

To avoid biasing the analysis, genuine data in the SR pass regions are not observed during their optimization.
Instead, the expected \rpf functions are derived through fits to the data in the VRs, and are then used in conjunction with the data in the SF regions to estimate the SP region background yields.
The VRs are used only to finalize the selections and validate the background estimation strategy.
They are not used in the search and do not affect the results presented in this paper.

\section{Systematic uncertainties}
\label{sec:SystUnc}

Several sources of systematic uncertainties in the signal and background estimation are considered and summarized in Table~\ref{tab:systs}.
The dominant source of uncertainty for both the fully merged and semimerged categories is the statistical uncertainty in the QCD multijet background estimate.
The impact of a nuisance parameter on a parameter of interest is defined as the shift in the parameter of interest when  the nuisance parameter is shifted from its best-fit value by its uncertainty, while profiling all other parameters~\cite{CMS:2024onh}.
The largest relative impact for a single transfer factor parameter, defined as the ratio of its impact on the best-fit signal strength to the total post-fit uncertainty in the signal strength, is 46\% and 28\% for the fully merged and semimerged categories, respectively, evaluated for the signal mass points with the largest global significances in each category.

In the fully merged category, this is followed by uncertainties in the signal efficiencies of the \particlenet \TXbb and \partmodel \TYWW selections, with relative impacts of 17 and 20\%, respectively.
The \TXbb selection efficiency scale factors are measured by comparing data and simulation in a data sample dominated by $\Pg\to\bbbar$ jets, obtained using the ``sfBDT'' method~\cite{CMS-BTV-22-001}, which uses a boosted decision tree to select $\Pg\to\bbbar$ jets that are similar to the resonant $\PH\to\bbbar$ signal.
The resulting scale factors range from 0.9 to 1.2 for the HP WP with uncertainties ranging from 4 to 10\%, depending on the jet \pt and data-taking era.
The dominant uncertainties in the scale factor include the effects of applying different assumptions for reweighting simulation to data as well as statistical uncertainties.
The \TYWW selection efficiency scale factors are evaluated using a new technique based on the ratio of the primary Lund jet plane~\cite{Dreyer:2018nbf, CMS:2025eyd, CMS-PAS-JME-25-001} densities of each individual subjet between data and simulation.
The scale factors for the latter range from 0.9 to 1.2 with uncertainties of ${\approx}10\%$ for highly boosted jets, \ie, from signals with a high $\ifrac{\mx}{\my}$ ratio, and range from 20 to 50\% for less boosted jets.
The dominant uncertainties in this method include an uncertainty stemming from corrections to simulations used to measure the Lund jet plane densities, as well as an uncertainty in the quark-to-subjet association.

In the semimerged category, the uncertainties in the \TXbb and \TWqq tagging efficiency have impacts of 13 and 9\%, respectively.
The same \TXbb selection efficiency scale factors are applied, while for the \TWqq selection, the efficiency is calibrated by comparing the efficiencies in data and simulation in a lepton plus jets data sample dominated by \ttbar events.
The scale factors for the \TWqq tagger WP used in the semimerged category range from 0.8 to 0.9 with uncertainties in the 5--10\% range depending on the jet \pt.

Other significant sources of experimental uncertainty include the scale and resolution of the regressed jet mass and reconstructed jet energy~\cite{CMS-DP-2021-033} in data versus simulation.
Jet mass corrections and uncertainties are measured in a data sample enriched in \ttbar events, using AK8 jets originating from hadronic \PW boson decays~\cite{CMS:2022gjd}, with impacts of 26 and 6\% in the fully merged and semimerged categories, respectively.
Jet energy corrections and uncertainties constitute 2 and 3\%, respectively, for the two categories.
Finally, there are large statistical uncertainties related to the MC simulations of the signal and less important \PQt and \PV backgrounds, reaching 30\% in some bins of the SRs.

Less important sources of uncertainty include those related to the renormalization and factorization scales, PDFs, Higgs boson branching fractions, integrated luminosity~\cite{CMS-LUM-17-003,CMS-PAS-LUM-20-001}, pileup effects, and trigger efficiencies, all of which have sub-percent-level impacts.

\begin{table}[htbp!]
    \caption{Sources of systematic uncertainties and the range of variations in the yields of the specified processes in the fully merged (FM) and semimerged (SM) categories.}
    \label{tab:systs}
    \resizebox*{\textwidth}{!}{
        \centering
        \begin{tabular}{lccc}
            Source                          & Processes affected & Uncertainty (FM)                & Uncertainty (SM) \\
            \hline
            \TYWW signal efficiency         & Signal             & 6--50\%                         & \NA              \\
            \TXbb signal efficiency         & Signal             & \multicolumn{2}{c}{$4$--$10\%$}                    \\
            \TWqq signal efficiency         & Signal             & \NA                             & 9\%              \\
            \hbb branching fraction         & Signal, \hbb       & \multicolumn{2}{c}{1.25\%}                         \\
            \rpf function parameters        & QCD multijet       & 8--20\%                         & 6\%              \\
            Parton showering                & All simulations    & \multicolumn{2}{c}{1--9\%}                         \\
            Jet energy scale and resolution & All simulations    & \multicolumn{2}{c}{1--3\%}                         \\
            Jet mass scale and resolution   & All simulations    & \multicolumn{2}{c}{1--4\%}                         \\
            MC statistical uncertainty      & Signal             & \multicolumn{2}{c}{1--10\%}                        \\
            MC statistical uncertainty      & \ttbar             & \multicolumn{2}{c}{1--30\%}                        \\
            Renorm. and fact. scales        & All simulations    & \multicolumn{2}{c}{7--10\%}                        \\
            PDF                             & All simulations    & \multicolumn{2}{c}{$1\%$}                          \\
            Integrated luminosity           & All simulations    & \multicolumn{2}{c}{$2\%$}                          \\
            Pileup                          & All simulations    & \multicolumn{2}{c}{$1\%$}                          \\
            Trigger efficiencies            & All simulations    & $<$1\%                          & 1--2\%           \\
        \end{tabular}
    }
\end{table}

\section{Results}
\label{sec:Results}

For both search categories, a 2D binned maximum likelihood fit is performed to the observed \mxrec and \myrec distributions, simultaneously in the fail and pass regions, for a wide range of potential \PX and \PY mass points.
The \mx and \my points are separated per category based on which one yields a lower expected exclusion limit; hence, the two categories themselves are not fit simultaneously.
The following results have been determined using the CMS statistical analysis tool \textsc{Combine}~\cite{CMS:2024onh}, based on the \textsc{RooFit}~\cite{Verkerke:2003ir} and \textsc{RooStats}~\cite{Moneta:2010pm} frameworks.

The data and fitted estimates for backgrounds and the signal hypotheses with the largest statistical significances for the fully merged and semimerged categories are shown in Figs.~\ref{fig:results_postfit_fm} and~\ref{fig:results_postfit_sr}, respectively, projected onto the \mxrec and \myrec observables.
Figure~\ref{fig:results_postfit_pulls2d} shows the 2D pull plots for both categories.
The largest deviation from the background-only hypothesis is observed in the fully merged category for $\mx = 900\GeV$ and $\my=80\GeV$ with a local (global) significance of 3.3 (below 1.0) standard deviations.
In the semimerged category, the largest deviation is observed for $\mx = 1200\GeV$ and $\my = 900\GeV$ with a local (global) significance of 1.7 (below 1.0) standard deviations.

Expected and observed exclusion limits at 95\% confidence level (\CL) are derived on the \XHY production cross section assuming a 100\% branching fraction for the \yww decays and standard model branching fractions for the subsequent \vvq decays.
They are shown in the 2D \mx vs. \my plane in Fig.~\ref{fig:results_xUL} as well as in one dimension for different \mx points in Fig.~\ref{fig:results_xUL_1D_ww}. The results can be interpreted in the context of \yzz decays by dividing the upper limits by a factor of 0.73 (0.78) in the fully merged (semimerged) category, to account for the lower selection efficiency of \yzz with respect to \yww, as explained in Section~\ref{sec:EventSelection}.
The separation between the fully merged and semimerged categories is governed by the fraction of signal \yvv jets containing four quarks and is depicted by a dashed line in Fig.~\ref{fig:results_xUL} and a separation between sets of points in Fig.~\ref{fig:results_xUL_1D_ww}. The loss of sensitivity at higher \my in Fig.~\ref{fig:results_xUL_1D_ww} corresponds to points where the decay products of the Higgs boson are no longer captured entirely within an AK8 jet and is similarly reflected by the increase in the limits along the boundary denoting kinematically allowed masses in Fig.~\ref{fig:results_xUL}.

The limits are calculated using asymptotic approximations for the profile likelihood ratio test statistic~\cite{ATLAS:2011tau, Cowan:2010js} and the \CLs~\cite{Junk:1999kv, Read:2002hq} criterion. The validity of the asymptotic approximation was tested for a subset of signals whose distributions lie predominantly in bins with less than ten expected events. For all such signals, the limits calculated using the asymptotic approximation agreed within 15\% with those calculated using the distribution of test statistics obtained from pseudodata, motivating the use of the asymptotic approximation for the full range of signal mass points.

The results were further interpreted in the context of asymmetric cascade decays in a two-real-scalar-singlet model~\cite{Robens:2020}, which introduces two real singlet fields \PS and \PX to the standard model Higgs sector.
Scans of the vacuum expectation values of these new scalars, $v_\PS$ and $v_\PX$, and the mixing angles between the Higgs field and the new fields as well as among themselves, $\theta_{\PH\PS}$, $\theta_{\PH\PX}$, and $\theta_{\PS\PX}$, are performed to obtain the maximally allowed product of cross section and branching fraction to \bbvv for all mass points under consideration. These scans employ the narrow width approximation to factorize production cross sections and branching fractions, consistent with the signal samples used in this analysis.
No mass exclusions can be made for any masses of the \PX and \PY bosons under this model, although for $\mx < 1.4$ and $\my < 0.6 \TeV$, the observed limits are about a factor of 10 higher than the maximum prediction of this model.

\begin{figure}[hbt]
    \centering

    \includegraphics[width=0.48\textwidth]{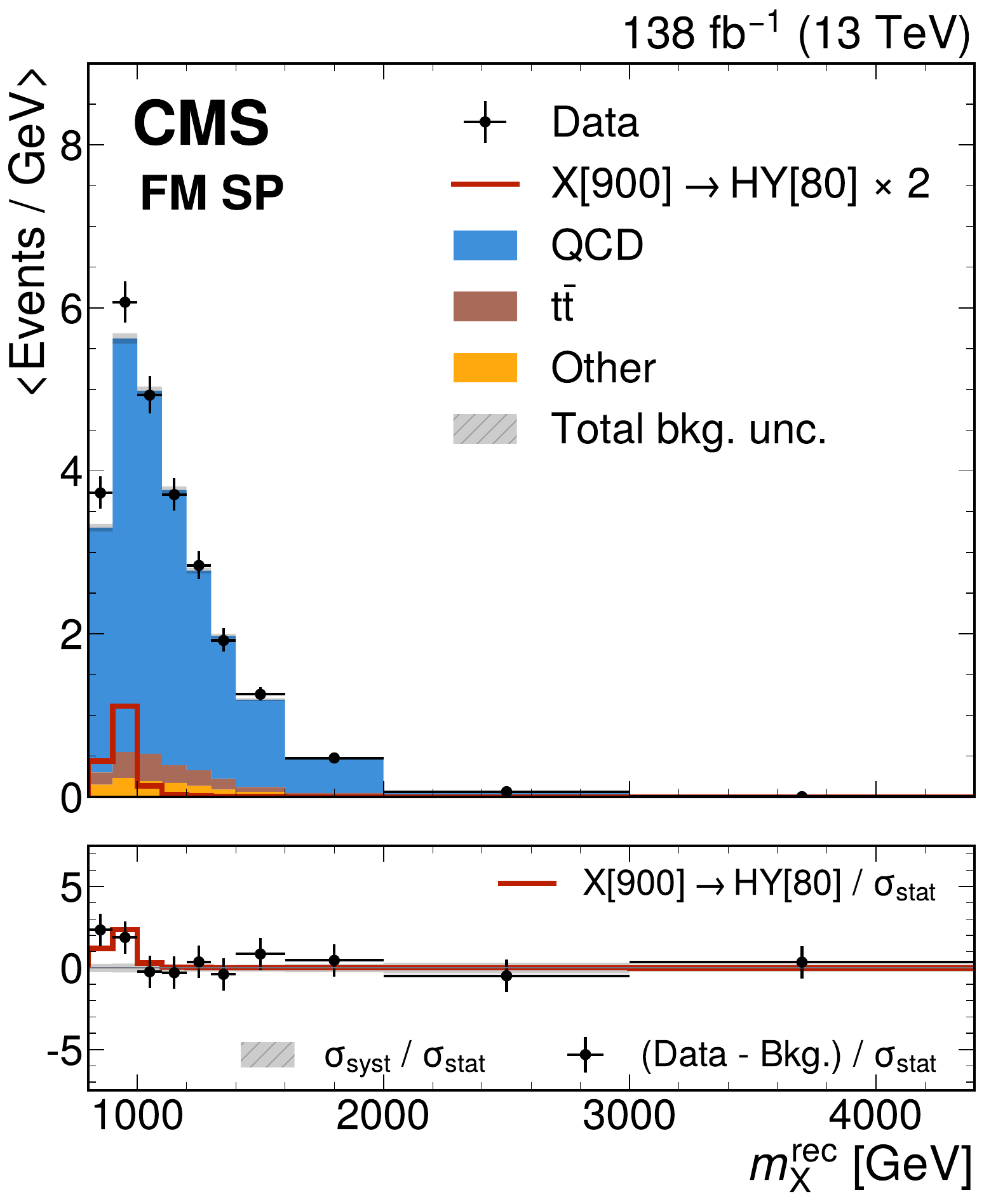}
    \includegraphics[width=0.51\textwidth]{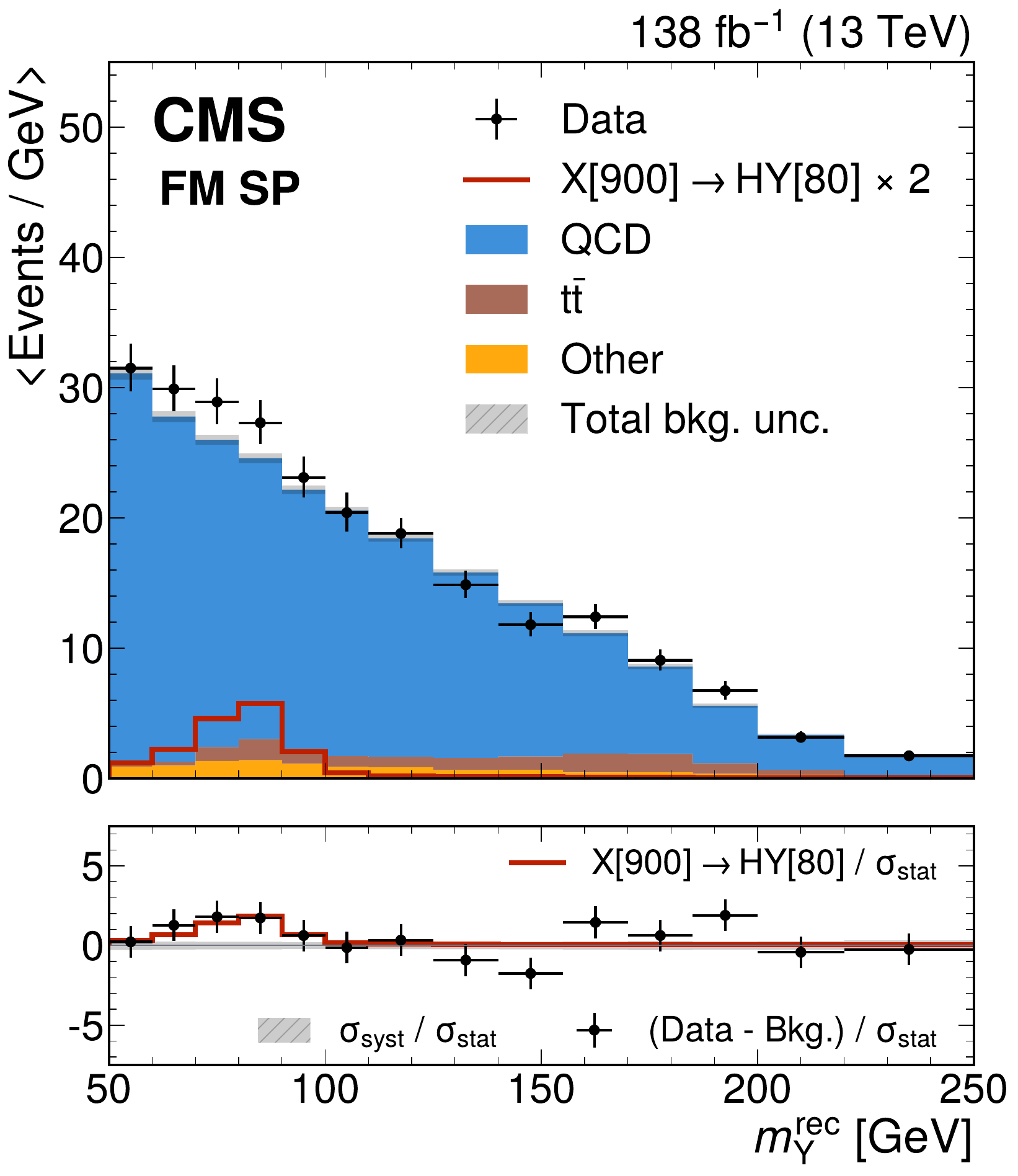}

    \caption{Projected distributions of the \mxrec (left) and \myrec (right) observables in the signal pass (SP) region of the fully merged (FM) category, after a maximum likelihood fit to the data with an \XHYbbWW signal with $\mx = 900\GeV$ and $\my = 80\GeV$ and product of best-fit production cross section and branching fraction of 15\unit{fb}.
        The signal distribution has been scaled by a factor of two in the upper panels to help distinguish it from the background distributions.
        The lower panels show the pull per bin, defined as $\ifrac{(\text{data} - \mathrm{Bkg.})}{\sstat}$, as well as a gray band indicating the systematic uncertainty in the pulls, estimated by dividing the corresponding systematic uncertainty in the main panel by $\sstat$.
        \label{fig:results_postfit_fm}
    }
\end{figure}

\begin{figure}[hbt]
    \centering

    \includegraphics[width=0.49\textwidth]{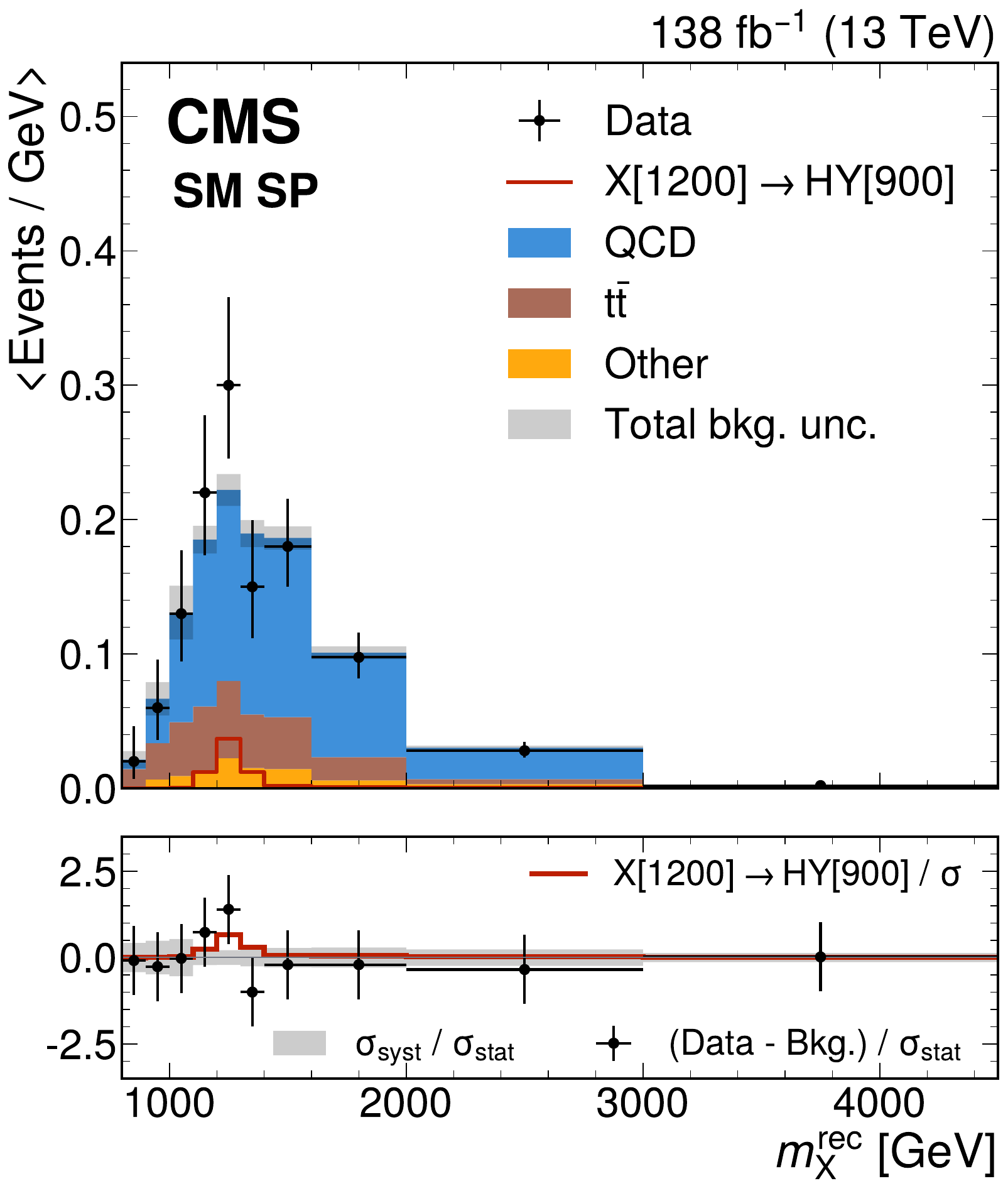}
    \includegraphics[width=0.49\textwidth]{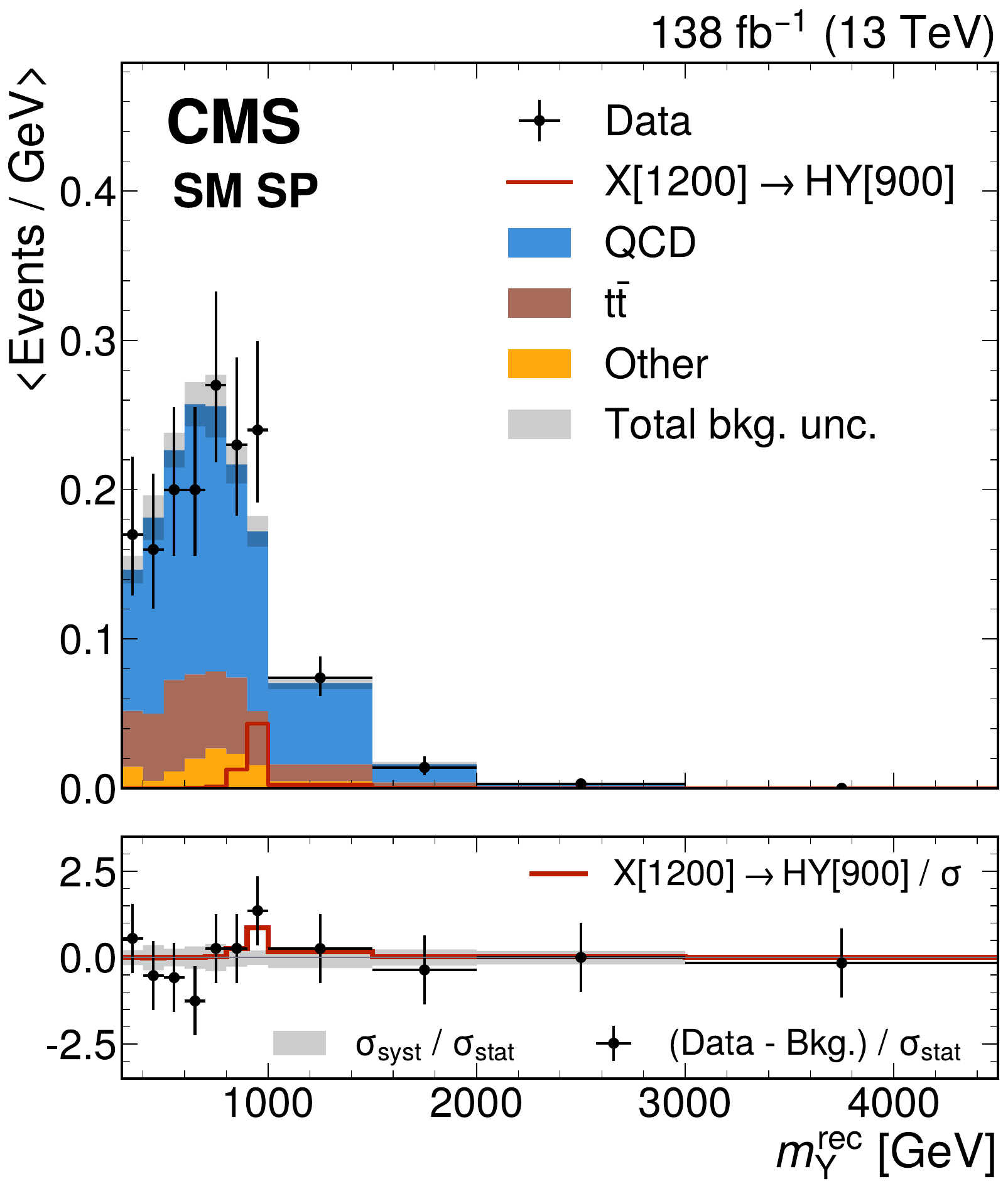}

    \caption{Projected distributions of the \mxrec (left) and \myrec (right) observables in the signal pass (SP) region of the semimerged (SM) category, after a maximum likelihood fit to the data with an \XHYbbWW signal with $\mx = 1200\GeV$ and $\my = 900\GeV$ and product of best-fit production cross section and branching fraction of 250\unit{fb}.
        The lower panels show the pull per bin, defined as $\ifrac{(\text{data} - \mathrm{Bkg.})}{\sstat}$, as well as a gray band indicating the systematic uncertainty in the pulls, estimated by dividing the corresponding systematic uncertainty in the main panel by $\sstat$.
        \label{fig:results_postfit_sr}
    }
\end{figure}

\begin{figure}[hbt]
    \centering

    \includegraphics[width=0.485\textwidth]{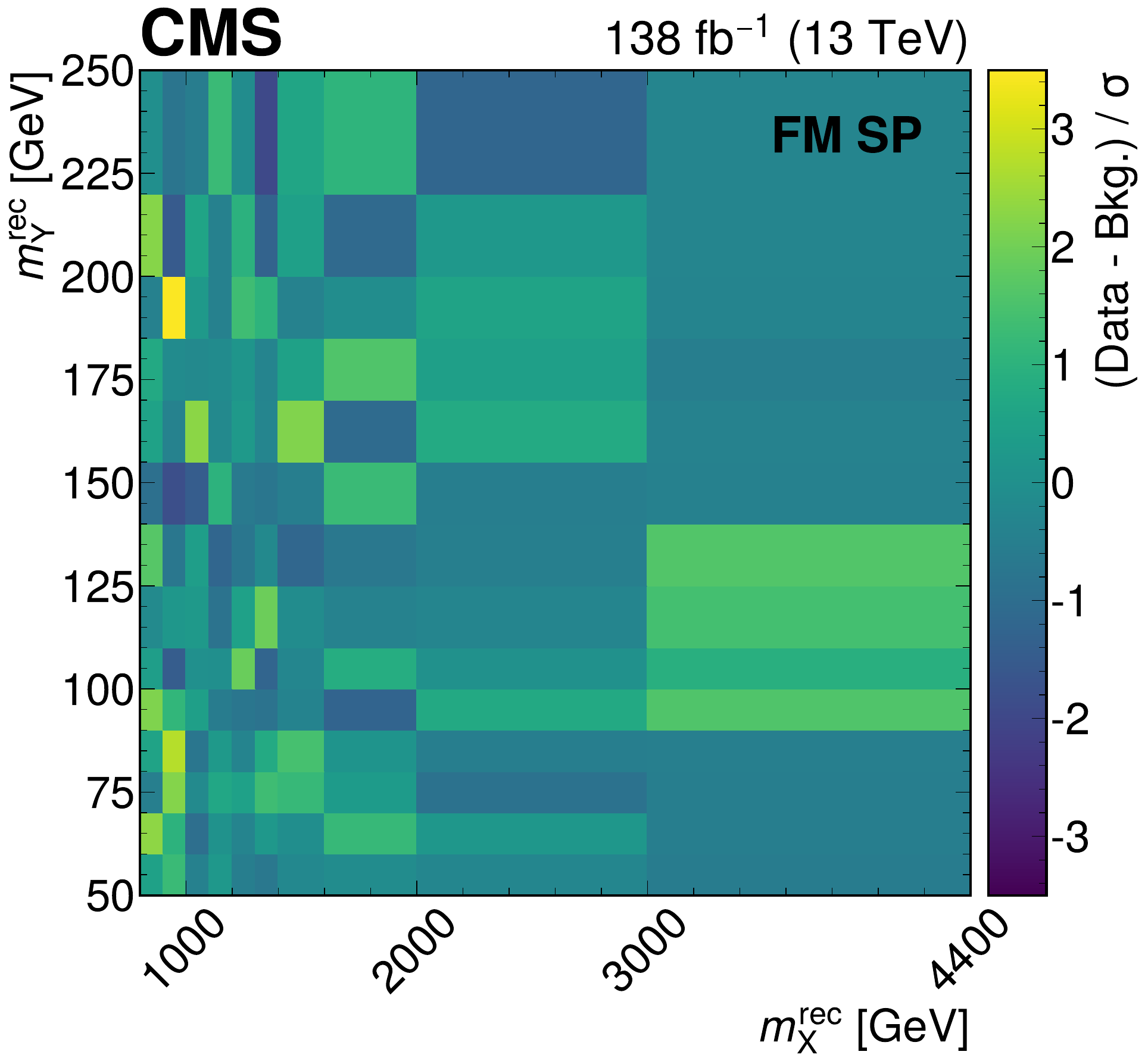}
    \includegraphics[width=0.485\textwidth]{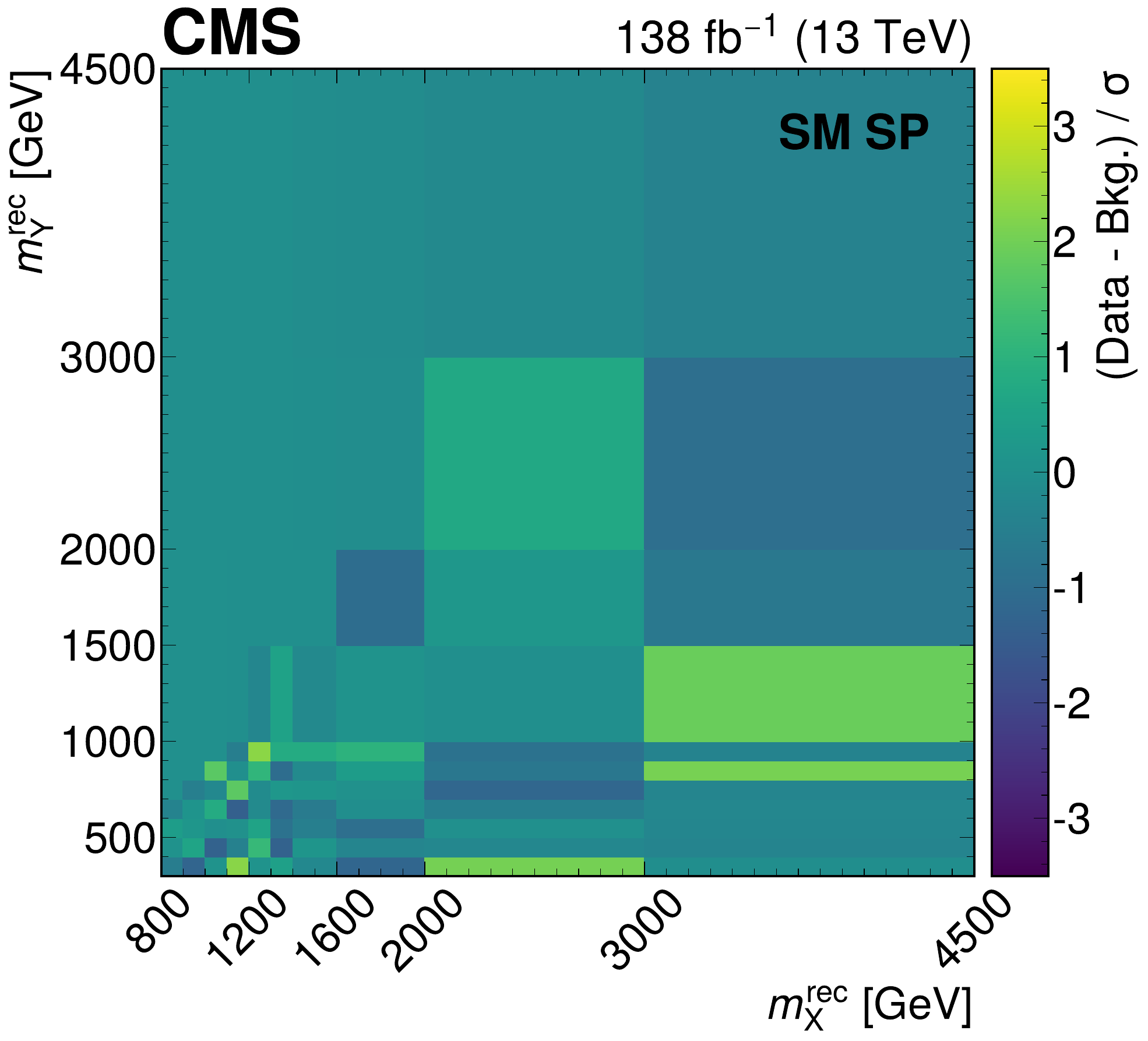}

    \caption{The 2D pulls in the \mxrec and \myrec observables in the signal pass (SP) regions of the fully merged (FM, left) and semimerged (SM, right) categories, after a maximum likelihood fit to the data with an \XHYbbWW signal with $\mx = 900\GeV$ and $\my = 80\GeV$, and $\mx = 1200\GeV$ and $\my = 900\GeV$, respectively.
        Pulls are defined as $\ifrac{(\text{data} - \mathrm{Bkg.})}{\sigma}$, where $\sigma$ is the sum in quadrature of the background uncertainty and the statistical uncertainty associated with the data assuming a Poisson distribution.
        \label{fig:results_postfit_pulls2d}
    }
\end{figure}

\begin{figure}[hbt]
    \centering

    \includegraphics[width=0.49\textwidth]{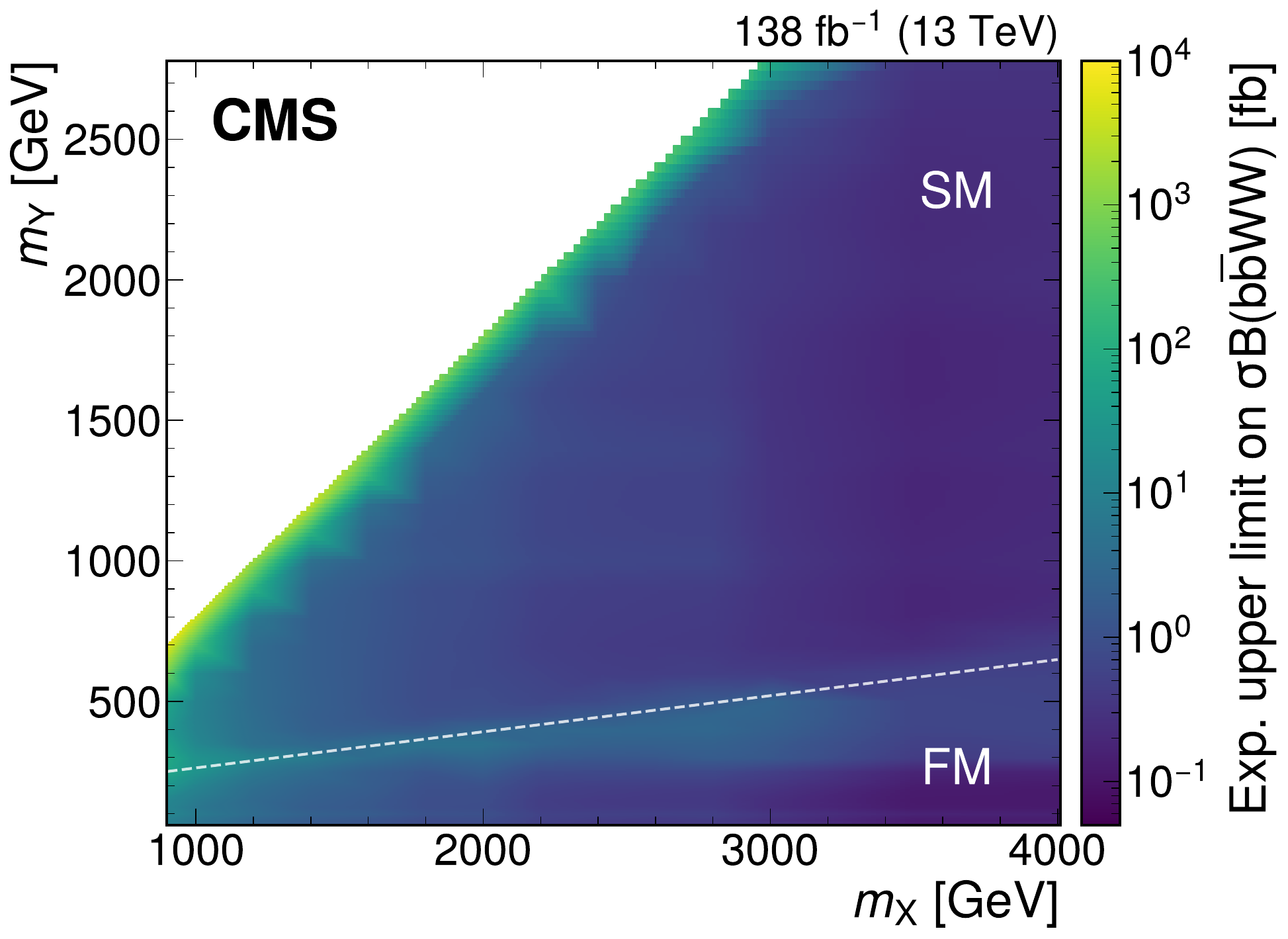}
    \includegraphics[width=0.49\textwidth]{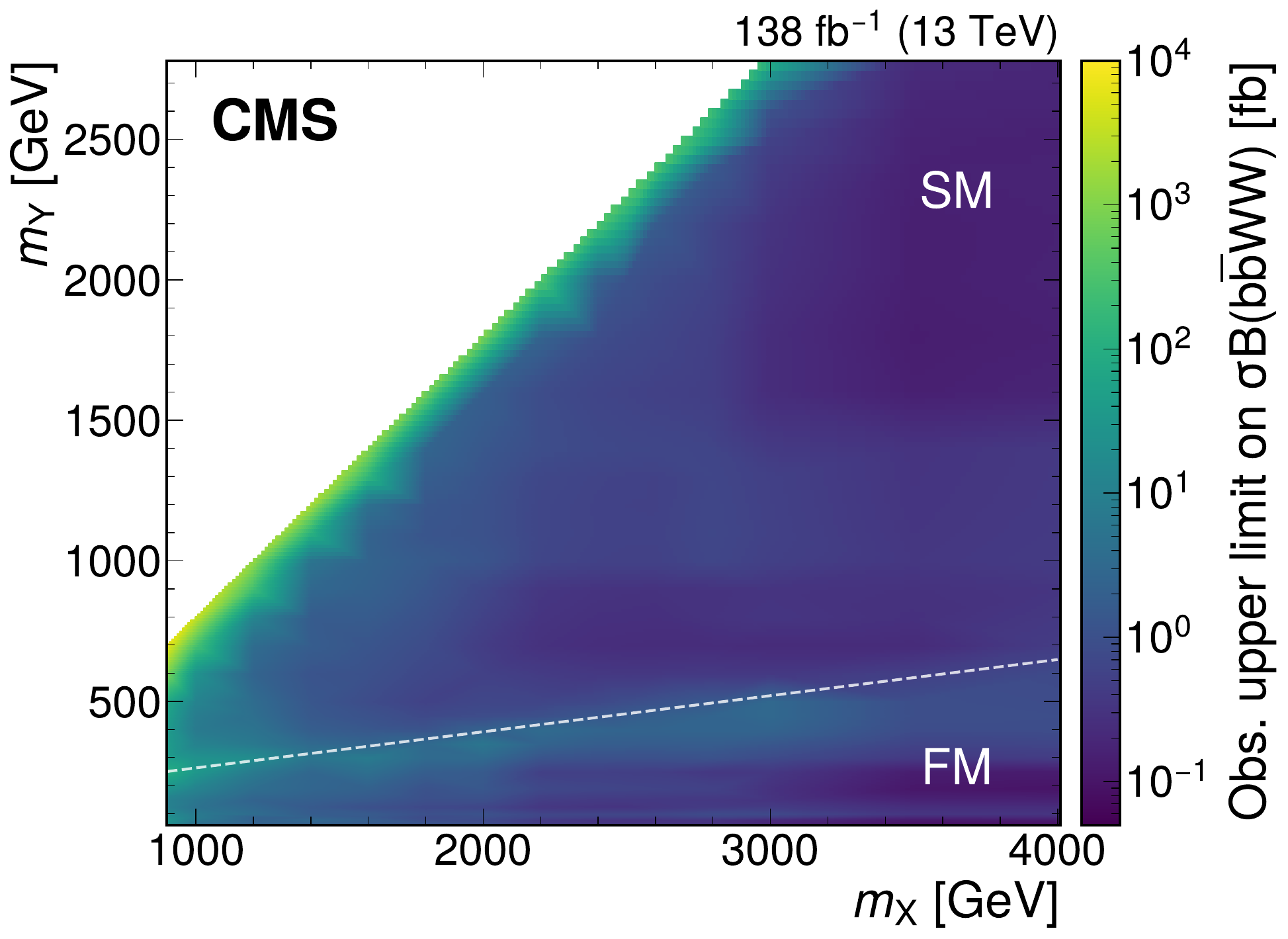}

    \caption{Median expected (left) and observed (right) upper limits at 95\% \CL on the product of production cross section and branching fraction for \XHYbbWW on the 2D \mx and \my plane.
        The dashed line separates the mass points considered in the fully merged (FM) and semimerged (SM) categories.
        \label{fig:results_xUL}
    }
\end{figure}

\begin{figure}[hbt]
    \centering

    \includegraphics[width=0.9\textwidth]{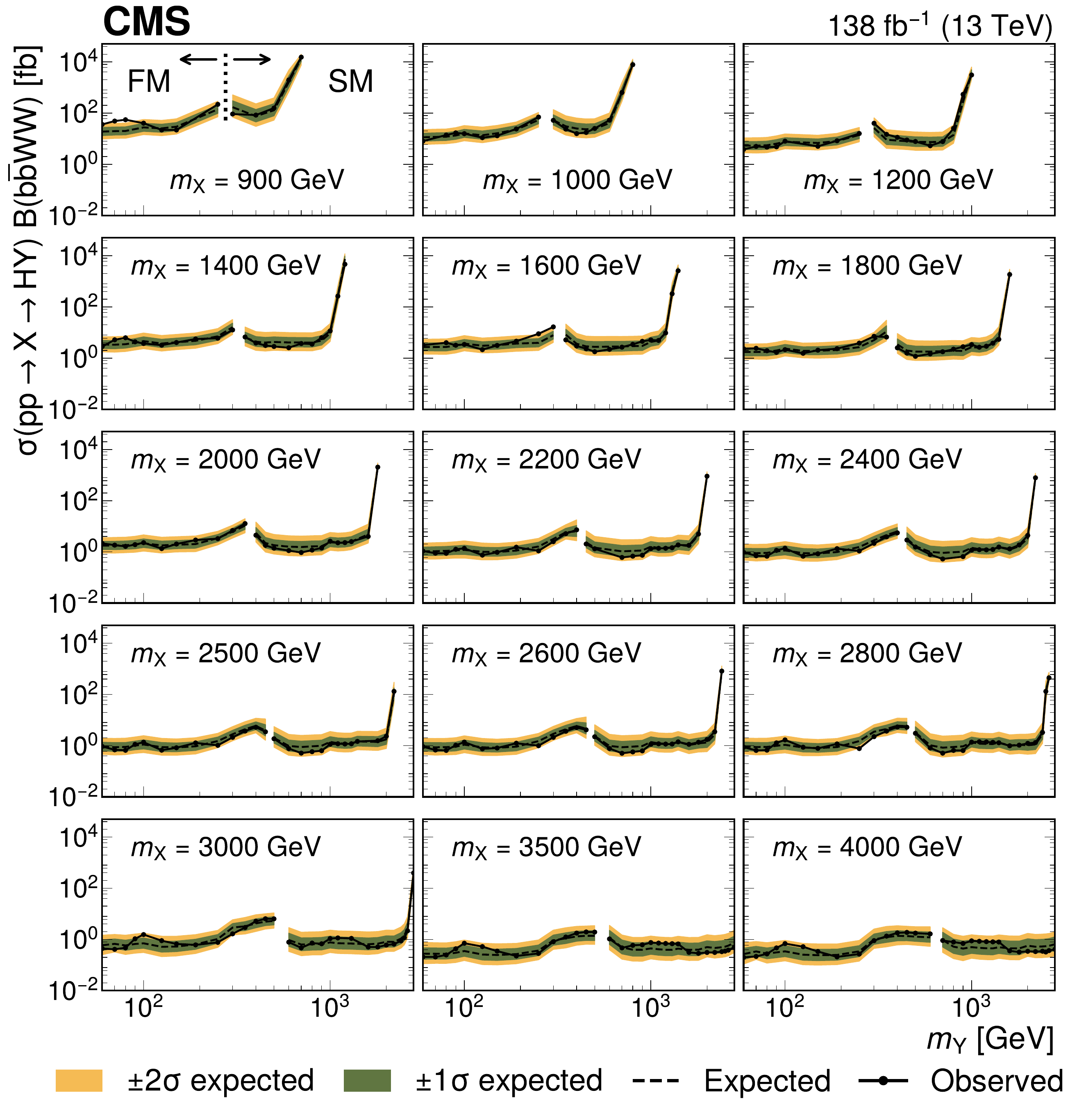}

    \caption{Expected and observed upper limits at 95\% \CL on the product of production cross section and branching fraction for \XHYbbWW signals.
        The dashed and solid black lines represent median expected and observed limits, respectively.
        The inner (green) and outer (yellow) bands represent the $\pm1$ and $\pm2$ standard deviations for the expected limits, respectively.
        The left and right set of points in each panel correspond to those considered in the fully merged (FM) and semimerged (SM) categories, respectively.
        \label{fig:results_xUL_1D_ww}
    }
\end{figure}

\clearpage
\section{Summary}
\label{sec:Summary}

A search has been performed for a heavy scalar resonance \PX decaying to a Higgs boson (\PH) and Higgs-like scalar boson (\PY) in the two bottom quark (\hbb) and four quark (\yvvq) final state, where \PV denotes a \PW or \PZ boson.
The search targets Lorentz-boosted \XHY production where the daughter quarks from the Higgs and \PY bosons are merged within large-area jets.
A mass-decorrelated graph neural network tagger, known as \particlenet, was used to select \hbb and \vqq jets while the new ``particle transformer'' is used for fully merged \yvv jets.
The signal is extracted from a two-dimensional maximum likelihood fit to data binned in the reconstructed \PY and \PX masses.
The dominant quantum chromodynamics multijet background is estimated using control regions in data with the tagger score selections inverted.
Other minor backgrounds including top quark and \PV plus jets are estimated using Monte Carlo simulations.
No significant excess is observed in the data above the standard model background expectation.
Upper limits at 95\% confidence level as low as 0.2\unit{fb} are set for the product of resonant \XHYbbVV production cross section and branching fraction for various \PX and \PY masses.
This is the first search for scalar resonances in the all-hadronic \bbvv final state at the LHC.

\begin{acknowledgments}
    We congratulate our colleagues in the CERN accelerator departments for the excellent performance of the LHC and thank the technical and administrative staffs at CERN and at other CMS institutes for their contributions to the success of the CMS effort. In addition, we gratefully acknowledge the computing centers and personnel of the Worldwide LHC Computing Grid and other centers for delivering so effectively the computing infrastructure essential to our analyses. Finally, we acknowledge the enduring support for the construction and operation of the LHC, the CMS detector, and the supporting computing infrastructure provided by the following funding agencies: SC (Armenia), BMBWF and FWF (Austria); FNRS and FWO (Belgium); CNPq, CAPES, FAPERJ, FAPERGS, and FAPESP (Brazil); MES and BNSF (Bulgaria); CERN; CAS, MoST, and NSFC (China); MINCIENCIAS (Colombia); MSES and CSF (Croatia); RIF (Cyprus); SENESCYT (Ecuador); ERC PRG, TARISTU24-TK10 and MoER TK202 (Estonia); Academy of Finland, MEC, and HIP (Finland); CEA and CNRS/IN2P3 (France); SRNSF (Georgia); BMFTR, DFG, and HGF (Germany); GSRI (Greece); NKFIH (Hungary); DAE and DST (India); IPM (Iran); SFI (Ireland); INFN (Italy); MSIT and NRF (Republic of Korea); MES (Latvia); LMTLT (Lithuania); MOE and UM (Malaysia); BUAP, CINVESTAV, CONACYT, LNS, SEP, and UASLP-FAI (Mexico); MOS (Montenegro); MBIE (New Zealand); PAEC (Pakistan); MES, NSC, and NAWA (Poland); FCT (Portugal);  MESTD (Serbia); MICIU/AEI and PCTI (Spain); MOSTR (Sri Lanka); Swiss Funding Agencies (Switzerland); MST (Taipei); MHESI (Thailand); TUBITAK and TENMAK (T\"{u}rkiye); NASU (Ukraine); STFC (United Kingdom); DOE and NSF (USA).

    \hyphenation{Rachada-pisek} Individuals have received support from the Marie-Curie program and the European Research Council and Horizon 2020 Grant, contract Nos.\ 675440, 724704, 752730, 758316, 765710, 824093, 101115353, 101002207, 101001205, and COST Action CA16108 (European Union); the Leventis Foundation; the Alfred P.\ Sloan Foundation; the Alexander von Humboldt Foundation; the Science Committee, project no. 22rl-037 (Armenia); the Fonds pour la Formation \`a la Recherche dans l'Industrie et dans l'Agriculture (FRIA) and Fonds voor Wetenschappelijk Onderzoek contract No. 1228724N (Belgium); the Beijing Municipal Science \& Technology Commission, No. Z191100007219010, the Fundamental Research Funds for the Central Universities, the Ministry of Science and Technology of China under Grant No. 2023YFA1605804, the Natural Science Foundation of China under Grant No. 12535004, and USTC Research Funds of the Double First-Class Initiative No.\ YD2030002017 (China); the Ministry of Education, Youth and Sports (MEYS) of the Czech Republic; the Shota Rustaveli National Science Foundation, grant FR-22-985 (Georgia); the Deutsche Forschungsgemeinschaft (DFG), among others, under Germany's Excellence Strategy -- EXC 2121 ``Quantum Universe" -- 390833306, and under project number 400140256 - GRK2497; the Hellenic Foundation for Research and Innovation (HFRI), Project Number 2288 (Greece); the Hungarian Academy of Sciences, the New National Excellence Program - \'UNKP, the NKFIH research grants K 131991, K 133046, K 138136, K 143460, K 143477, K 146913, K 146914, K 147048, 2020-2.2.1-ED-2021-00181, TKP2021-NKTA-64, and 2025-1.1.5-NEMZ\_KI-2025-00004 (Hungary); the Council of Science and Industrial Research, India; ICSC -- National Research Center for High Performance Computing, Big Data and Quantum Computing, FAIR -- Future Artificial Intelligence Research, and CUP I53D23001070006 (Mission 4 Component 1), funded by the NextGenerationEU program (Italy); the Latvian Council of Science; the Ministry of Education and Science, project no. 2022/WK/14, and the National Science Center, contracts Opus 2021/41/B/ST2/01369, 2021/43/B/ST2/01552, 2023/49/B/ST2/03273, and the NAWA contract BPN/PPO/2021/1/00011 (Poland); the Funda\c{c}\~ao para a Ci\^encia e a Tecnologia, grant CEECIND/01334/2018 (Portugal); the National Priorities Research Program by Qatar National Research Fund;  MICIU/AEI/10.13039/501100011033, ERDF/EU, ``European Union NextGenerationEU/PRTR", and Programa Severo Ochoa del Principado de Asturias (Spain); the Chulalongkorn Academic into Its 2nd Century Project Advancement Project, the National Science, Research and Innovation Fund program IND\_FF\_68\_369\_2300\_097, and the Program Management Unit for Human Resources \& Institutional Development, Research and Innovation, grant B39G680009 (Thailand); the Kavli Foundation; the Nvidia Corporation; the SuperMicro Corporation; the Welch Foundation, contract C-1845; and the Weston Havens Foundation (USA).
\end{acknowledgments}\section*{Data availability} Release and preservation of data used by the CMS Collaboration as the basis for publications is guided by the  \href{https://doi.org/10.7483/OPENDATA.CMS.1BNU.8V1W}{CMS data preservation, re-use and open access policy}.

\bibliography{auto_generated}

\cleardoublepage \appendix\section{The CMS Collaboration \label{app:collab}}\begin{sloppypar}\hyphenpenalty=5000\widowpenalty=500\clubpenalty=5000\cmsinstitute{Yerevan Physics Institute, Yerevan, Armenia}
{\tolerance=6000
A.~Hayrapetyan, V.~Makarenko\cmsorcid{0000-0002-8406-8605}, A.~Tumasyan\cmsAuthorMark{1}\cmsorcid{0009-0000-0684-6742}
\par}
\cmsinstitute{Institut f\"{u}r Hochenergiephysik, Vienna, Austria}
{\tolerance=6000
W.~Adam\cmsorcid{0000-0001-9099-4341}, J.W.~Andrejkovic, L.~Benato\cmsorcid{0000-0001-5135-7489}, T.~Bergauer\cmsorcid{0000-0002-5786-0293}, M.~Dragicevic\cmsorcid{0000-0003-1967-6783}, C.~Giordano, P.S.~Hussain\cmsorcid{0000-0002-4825-5278}, M.~Jeitler\cmsAuthorMark{2}\cmsorcid{0000-0002-5141-9560}, N.~Krammer\cmsorcid{0000-0002-0548-0985}, A.~Li\cmsorcid{0000-0002-4547-116X}, D.~Liko\cmsorcid{0000-0002-3380-473X}, M.~Matthewman, I.~Mikulec\cmsorcid{0000-0003-0385-2746}, J.~Schieck\cmsAuthorMark{2}\cmsorcid{0000-0002-1058-8093}, R.~Sch\"{o}fbeck\cmsAuthorMark{2}\cmsorcid{0000-0002-2332-8784}, D.~Schwarz\cmsorcid{0000-0002-3821-7331}, M.~Shooshtari\cmsorcid{0009-0004-8882-4887}, M.~Sonawane\cmsorcid{0000-0003-0510-7010}, W.~Waltenberger\cmsorcid{0000-0002-6215-7228}, C.-E.~Wulz\cmsAuthorMark{2}\cmsorcid{0000-0001-9226-5812}
% ...
\par}
\vskip\cmsinstskip
\dag:~Deceased\\
$^{1}$Also at Yerevan State University, Yerevan, Armenia\\
$^{2}$Also at TU Wien, Vienna, Austria\\
% ...
\end{sloppypar}
\end{document}